\documentclass[
showpacs,preprintnumbers,amsmath,amssymb,nofootinbib]{revtex4}

\usepackage{amsmath} \usepackage{graphicx} \usepackage{amsfonts}
\usepackage{array} \usepackage{amsthm}

\usepackage{latexsym}
\usepackage{epic}
\usepackage{eepic}
\usepackage{graphicx,psfrag}

\newcommand{\e}{{\rm e}}
\newcommand{\ds}{\displaystyle}

\newcommand{\de}{\delta}

\newcommand{\bw}{\begin{widetext}}
\newcommand{\ew}{\end{widetext}}
\newcommand{\be}{\begin{equation}}
\newcommand{\ee}{\end{equation}}
\newcommand{\bea}{\begin{eqnarray}}
\newcommand{\eea}{\end{eqnarray}}

\newcommand{\fns}{\footnotesize}
\newcommand{\scs}{\scriptsize}

\begin{document}


\title{Extremal black holes in D=4 Gauss-Bonnet gravity}

\author{Chiang-Mei Chen} \email{cmchen@phy.ncu.edu.tw}
\affiliation{Department of Physics, National Central University,
Chungli 320, Taiwan}

\author{Dmitri V. Gal'tsov} \email{galtsov@phys.msu.ru}
\affiliation{Department of Theoretical Physics, Moscow State
University, 119899, Moscow, Russia}

\author{Dmitry G. Orlov} \email{orlov_d@mail.ru}

\affiliation{Department of Physics, National Central University,
Chungli 320, Taiwan}

\affiliation{Center for Gravitation and Fundamental Metrology,
VNIIMS, \\ 46 Ozyornaya St., Moscow 119361, Russia}

\date{\today}

\begin{abstract}
We show that four-dimensional Einstein-Maxwell-Dilaton-Gauss-Bonnet
gravity admits asymptotically flat black hole solutions with a
degenerate event horizon of the Reissner-Nordstr\"om type
$AdS_2\times S^2$. Such black holes exist for the dilaton coupling
constant within the interval $0\leq a^2<a^2_{\rm cr}$. Black holes
must be endowed with an electric charge and (possibly) with magnetic
charge (dyons) but they can not be purely magnetic. Purely electric
solutions are constructed numerically and the critical dilaton
coupling is determined $a_{\rm cr}\simeq 0.488219703$. For each
value of the dilaton coupling $a$ within this interval and for a
fixed value of the Gauss--Bonnet coupling $\alpha$ we have a family
of black holes parameterized by their electric charge. Relation
between the mass, the electric charge and the dilaton charge at both
ends of the allowed interval of $a$  is reminiscent of the BPS
condition for dilaton black holes in the Einstein-Maxwell-Dilaton
theory. The entropy of the DGB extremal black holes is twice the
Bekenstein-Hawking entropy.
\end{abstract}

\pacs{04.20.Jb, 04.65.+e, 98.80.-k}

\maketitle

\section{Introduction}
String theory suggests higher curvature corrections to the Einstein
equations \cite{Zwiebach:1985uq, Callan:1986jb, Gross:1986iv}. Black
holes in higher-curvature gravity were extensively studied during
two past  decades \cite{Myers:1998gt, Callan:1988hs} culminating in
recent spectacular progress in the microscopic string calculations
of the black hole entropy (for a review see \cite{deWit:2005ya,
Mohaupt:2005jd}). In theories with higher curvature corrections,
classical entropy deviates from the Bekenstein-Hawking value and can
be calculated using Wald's formalism \cite{Wald:1993nt,
Jacobson:1993vj, Iyer:1994ys, Jacobson:1994qe}. Remarkably, it still
exhibits exact agreement with string theory quantum predictions at
the corresponding level, both in the BPS \cite{Behrndt:1998eq,
LopesCardoso:1998wt, LopesCardoso:1999cv, LopesCardoso:1999ur,
LopesCardoso:1999xn, Mohaupt:2000mj, LopesCardoso:2000qm,
LopesCardoso:2000fp, Dabholkar:2004yr, Dabholkar:2004dq, Sen:2004dp,
Hubeny:2004ji, Bak:2005mt} and non-BPS \cite{Goldstein:2005hq,
Kallosh:2005ax, Tripathy:2005qp, Giryavets:2005nf, Goldstein:2005rr,
Kallosh:2006bt, Kallosh:2006bx, Prester:2005qs, Alishahiha:2006ke,
Sinha:2006yy, Chandrasekhar:2006kx, Parvizi:2006uz, Sahoo:2006rp,
Astefanesei:2006sy} cases. In some supersymmetric models with higher
curvature terms exact classical solutions for static black holes
were obtained \cite{Dabholkar:2004yr, Dabholkar:2004dq, Bak:2005mt}.
Moreover, as was argued by Sen \cite{Sen:2005wa, Sen:2005iz},
knowledge of exact global black hole solutions is not necessary to
be able to compare classical and quantum results: the entropy can be
computed locally using  the ``entropy function'' approach based on
the typical for supergravities attractor property
\cite{Goldstein:2005hq, Kallosh:2005ax, Tripathy:2005qp,
Giryavets:2005nf, Goldstein:2005rr, Kallosh:2006bt}. In this case it
is tacitly assumed that global asymptotically flat black hole
solutions exist indeed. Generically, however, the existence of local
solutions does not imply possibility to extend them to infinity as
asymptotically flat black holes. Here we investigate this issue
within a simple model of higher curvature gravity.

One of the simplest four-dimensional models with higher-curvature
terms is the so-called dilaton-Gauss-Bonnet gravity (DGB)  which is
obtained by adding to the Einstein action the four-dimensional Euler
density multiplied by the dilaton exponent. As other
higher-curvature theories based on topological invariants, this
theory does not contain higher derivatives and thus is free of
ghosts. Black hole solutions in this theory can not be found in
analytical form, but they were extensively studied perturbatively
\cite{Mignemi:1992nt, Mignemi:1993ce} and numerically
\cite{Kanti:1995vq, Torii:1996yi, Alexeev:1996vs, Alexeev:1997ua}.
More recently global properties of DGB black hole solutions were
studied using the dynamical system approach
\cite{Melis:2005xt,Melis:2005ji,Mignemi:2006,Melis:2006}. Stability
issues were discussed in
\cite{Torii:1998gm,Dotti:2004sh,Dotti:2005sq,Gleiser:2005ra,Moura:2006pz}.
In these papers the existence of both neutral and charged
asymptotically flat solutions with a non-degenerate event horizon
and without naked singularities was established. These solutions
have the Schwarzschild type event horizon and do not possess an
extremal limit. In this respect they differs substantially from the
dilatonic black holes in the Einstein-Maxwell-dilaton (EMD) theory
without the GB term
\cite{Gibbons:1982ih,Gibbons:1985ac,Gibbons:1987ps,Garfinkle:1990qj}:
charged dilatonic black holes do have an extremal limit in which
case the event horizon shrinks to a point-like singularity. The
Bekenstein-Hawking entropy of the extremal dilatonic black holes is
zero, while quantum theory suggests a non-zero result. The puzzle
was solved in several supersymmetric models by showing  that the
horizon of the extremal dilatonic black hole is stretched to a
finite radius. In the case of the DGB black holes, however, no
solution  with the {\em degenerate} event horizon of finite radius
was found so far.

The aim of the present paper is to study this possibility in more
detail. We show that, apart from the  known DGB black holes with a
non-degenerate event horizon, there exist electrically charged
solutions with the degenerate horizon of the $AdS_2 \times S^2$ type
which are asymptotically flat. These new solutions exist only in a
limited range of the dilaton coupling constant. For other values of
this  constant, local solutions with the $AdS_2 \times S^2$  horizon
can not be continued to infinity as asymptotically flat black holes:
singularity is met in a finite distance outside the horizon. Since
the DGB theory does not possess S-duality, magnetic solutions differ
substantially from the electric ones; in particular, no purely
magnetic black holes with a degenerate horizon are allowed, though
dyonic solutions with a non-zero electric charge are possible.

\section{4D dilatonic Gauss-Bonnet theory}
The action for the four-dimensional Einstein-Maxwell-dilaton
theory with an arbitrary dilaton coupling constant $a$ modified by
the DGB term reads
\begin{equation}\label{reducedS}
S = \frac1{16\pi} \int  \left\{ R - 2 \partial_\mu \phi
\partial^\mu \phi - {\rm e}^{2 a \phi} \left(F^2 - \alpha {\cal
L}_{\rm GB} \right) \right\}\sqrt{-g}\,d^4 x ,
\end{equation}
where ${\cal L}_{\rm GB}$ is the Euler density
\begin{equation}
{\cal L}_{\rm GB} = R^2 - 4 R_{\mu\nu} R^{\mu\nu} +
R_{\alpha\beta\mu\nu} R^{\alpha\beta\mu\nu}.
\end{equation}
This action contains two parameters (we use units $G=c=1$): the
dilaton coupling $a$ and the GB coupling $\alpha$. We will assume $a
\geq 0,\; \alpha \geq 0$. Solutions for negative $a$ can be obtained
changing the sign of the dilaton. Note that the Maxwell term is not
multiplied by $\alpha$ to facilitate decoupling of the GB term from
the EMD action.
\par
Consider the static spherically symmetric metrics parameterized for
further convenience by three functions $w(r),\, \rho(r),\,
\sigma(r)$:
\begin{equation}
ds^2 = - w(r)\sigma^2(r) dt^2 + \frac{dr^2}{w(r)} + \rho^2(r)
d\Omega_2^2,
\end{equation}
The scalar curvature and the
Euler density then read
\begin{eqnarray}
R &=& \frac1{\sigma \rho^2} \left\{ - (4 \sigma w \rho \rho' +
\sigma w' \rho^2 + 2 \sigma' w \rho^2)' + 2 \sigma [\rho' (w \rho)'
+ 1] + 4 \sigma' w \rho \rho' \right\},
\\
{\cal L}_{\rm GB} &=& \frac4{\sigma \rho^2} \left[
\frac{(w\sigma^2)' (w\rho'^2 - 1)}{\sigma} \right]'.
\end{eqnarray}
The corresponding ansatz for the Maxwell one-form is
\begin{equation}
A = - f(r) \, dt - q_m \cos\theta \, d\varphi,
\end{equation}
where $f(r)$ is the electrostatic potential and $q_m$ is the
magnetic charge. Note that the DGB term breaks the discrete
S-duality   which in absence of this term is described by the
transformation
\begin{equation}
g_{\mu\nu} \to g_{\mu\nu}, \quad F \to {\rm e}^{-2a\phi} {}^* F,
\quad \phi \to - \phi,
\end{equation}
where $F=dA$. It is expected therefore that properties of electric
or magnetic black holes in this theory will be be essentially
different.

\subsection{Reduced action and field equations}
The corresponding one-dimensional Lagrangian is obtained by
dropping the total derivative in the dimensionally reduced action:
\begin{equation}
L = \frac{\sigma}2  [\rho' (w \rho)' + 1] + \sigma' w \rho \rho' - 2
\alpha a \sigma^{-1} (\sigma^2 w)' (w \rho'^2 - 1) \phi' {\rm e}^{2
a \phi}  - \frac{\sigma}2 w \rho^2 \phi'^2 + \frac{1}2
\frac{\rho^2}{\sigma} f'^2 {\rm e}^{2 a \phi} - \frac{1}2
\frac{\sigma}{\rho^2} q_m^2 {\rm e}^{2 a \phi}.
\end{equation}
The corresponding equations of motion read:
\begin{eqnarray}
8 \alpha a \left[ w (w \rho'^2 - 1) \phi' {\rm e}^{2 a \phi}
\right]' - \rho' (w \rho)' + 1 - 2 w \rho \rho'' - 4 \alpha a w'
(w \rho'^2 - 1) \phi' {\rm e}^{2 a \phi} - w \rho^2 \phi'^2 &&
\nonumber\\
- \frac{ \rho^2}{\sigma^2} f'^2 {\rm e}^{2 a \phi} - \frac{
q_m^2}{\rho^2} {\rm e}^{2 a \phi} &=& 0, \label{Eqf1}
\\
4 \alpha a \left[ \frac{(w \rho'^2 - 1) \phi' {\rm e}^{2 a
\phi}}{\sigma} \right]' \sigma - 4 \alpha a \frac{(\sigma^2
w)'}{\sigma^2} \rho'^2 \phi' {\rm e}^{2 a \phi} - \rho \rho'' +
\rho \rho' \frac{\sigma'}{\sigma} - \rho^2 \phi'^2 &=& 0,
\label{Eqw}
\\
4 \alpha a \left[ \frac{w \rho' (\sigma^2 w)' \phi' {\rm e}^{2 a
\phi}}{\sigma} \right]' - \frac12 \left[ \frac{(\sigma^2
w)'}{\sigma} \right]' \rho - (\sigma w \rho')' - \sigma w \rho
\phi'^2 + \frac{ \rho}{\sigma} f'^2 {\rm e}^{2 a \phi} + \sigma
\frac{  q_m^2}{\rho^3} {\rm e}^{2 a \phi} &=& 0, \label{Eqrho}
\\
(\sigma w \rho^2 \phi')' + 2 \alpha a \left[ \frac{(\sigma^2 w)' (w
\rho'^2 - 1)}{\sigma} \right]' {\rm e}^{2 a \phi} + a
\frac{\rho^2}{\sigma} f'^2 {\rm e}^{2 a \phi} - a  \sigma
\frac{q_m^2}{\rho^2} {\rm e}^{2 a \phi} &=& 0, \label{Eqphi}
\\
\left( \frac{\rho^2}{\sigma} f' {\rm e}^{2 a \phi} \right)' &=& 0.
\label{Eqf}
\end{eqnarray}
The Maxwell equation for the form field (\ref{Eqf}) can be directly
solved
\begin{equation}\label{Solf}
f'(r) = q_e \sigma \rho^{-2} {\rm e}^{- 2 a \phi},
\end{equation}
where $q_e$ is the electric charge.

\subsection{Global symmetries and conserved quantities}
The action (\ref{reducedS}) is invariant under the following
three-parametric group of global transformations:
\begin{equation}\label{symL}
w \to w \, {\rm e}^{-2 (\delta + \lambda)}, \quad \rho \to \rho \,
{\rm e}^{\delta}, \quad r \to r \, {\rm e}^{- \lambda} + \nu,
\quad \sigma \to \sigma \, {\rm e}^{\lambda}, \quad \phi \to \phi
+ \frac{\delta}{a}, \quad f \to f \, {\rm e}^{-2 \delta}.
\end{equation}
They generate three conserved Noether currents
\begin{equation}
J_g := \left( \frac{\partial L}{\partial \Phi'^A} \Phi'^A - L
\right) \partial_g r \bigg|_{g=0} - \frac{\partial L}{\partial
\Phi'^A} \, \partial_g \Phi^A \bigg|_{g=0},\quad \partial_r J_g=0,
\end{equation}
where $\Phi^A$ stands $\sigma, w , \rho, \phi$, $f$, and $g =
\delta,\, \nu,\, \lambda$. The conserved quantity corresponding to
the parameter $\nu$ is the  Hamiltonian
\begin{equation}\label{J1}
H = \frac12 \sigma [\rho' (w \rho)' - 1] + \sigma' w \rho \rho'  - 2
\alpha a \sigma^{-1} (\sigma^2 w)' (3 w \rho'^2 - 1) \phi' {\rm
e}^{2 a \phi} - \frac12 \sigma w \rho^2 \phi'^2 + \frac{1}2
\frac{\rho^2}{\sigma} f'^2 {\rm e}^{2 a \phi} + \frac{1}2
\frac{\sigma}{\rho^2} q_m^2 {\rm e}^{2 a \phi}.
\end{equation}
This is known to vanish on shell for diffeomorphism invariant
theories, $H = 0$.   The Noether current corresponding to the
parameter $\delta$ leads to the conservation equation $J_\de=0$,
where
\begin{equation}
J_\de= \frac{\sigma w \rho^2 \phi'}a  - \frac{(\sigma^2 w)'
\rho^2}{2\sigma} + 2 q_e f  + 2 \alpha \Bigl( \sigma^{-1}
\left\{ (w \rho'^2 - 1) \left[ (\sigma^2 w)' - 2 a \sigma^2 w \phi'
\right] + 2 a w (\sigma^2 w)' \rho \rho' \phi' \right\} {\rm e}^{2 a
\phi} \Bigr),
\end{equation}
which is an Abelian counterpart of the equation given in
\cite{Donets:1995ya}\footnote{We use this occasion to correct a
misprint in Ref. \cite{Donets:1995ya}: the factor
$\mathrm{e}^{-2\Phi}$ is missing at the right hand side of the
Eq.(23).}. The value of this integral depends on solutions. The
third integral corresponding to $\lambda$ is trivial: $J_\lambda = -
r H$, its existence implies $H = 0$.
\par
The above integrals of motion allows one to reduce the order of the
system by two. Fixing the gauge (e.g. $\sigma=1$) one has three
second order equations for $w,\,\rho,\,\phi$ with $q_e,\, q_m$
entering as parameters of this six-order system. Using the
integrals, the system order can reduced to four, with one parameter
more (the fixed value of $J_\de$). For $q_e=0$ one can further
reduce the order to three. Introducing, for instance, new variables
\begin{equation}
w \to \exp(w), \quad \rho \to \exp \left(\rho - \frac{w}2 \right),
\quad \phi \to \phi - \frac1{2a}w,
\end{equation}
we exclude from the system the variable $w$ (while $w'$ and $w''$
still persist). For numerical computations we use the initial
six-dimensional system, checking the constancy of the integrals of
motion to control accuracy of the calculation.
\par
The space of solutions is invariant under a {\em four-parameteric}
group of global transformations which  consists in rescaling of the
electric charge
\begin{equation}
q_e \to q_e \, {\rm e}^{2 \delta}, \qquad q_m \to q_m,
\end{equation}
(leaving the magnetic charge invariant), rescaling and shift of an
independent variable
\begin{equation}\label{trr}
r \to r \, {\rm e}^{\frac{\mu}2 + \delta} + \nu,
\end{equation}
and the following transformation of the field functions:
\begin{equation} \label{symsol}
w \to w \, {\rm e}^{\mu}, \quad \rho \to \rho \, {\rm e}^{\delta},
\quad \sigma \to \sigma \, {\rm e}^{\lambda}, \quad \phi \to \phi +
\frac{\delta}{a}, \quad f \to f \, {\rm e}^{\frac{\mu}2 - \delta +
\lambda}.
\end{equation}
Note that the Lagrangian is rescaled under this 4-parameter
transformations as ${\sl L} \to {\rm e}^{\lambda} {\sl L} $, so  the
action (\ref{reducedS}) remains invariant provided
\begin{equation}
\mu = -2 (\delta + \lambda),
\end{equation}
in which case we go back to (\ref{symL}). The shift $\nu$ is trivial
and the symmetry related to $\lambda$ can be frozen by the gauge
choice $\sigma = 1$. Therefore, physically interesting
transformations are generated by  $\mu$ and $\delta$ forming the
subgroup which we denote as $G(\mu, \delta)$.

\subsection{Turning points of $\rho(r)$ and the gauge choice}
Re-parametrization of the radial variable $r$ allows to eliminate
one of the three metric functions. There are two convenient gauge
choices: the Schwarzschild gauge $\rho = \bar r$, in which the
radial variable is the radius of two-spheres:
\begin{equation}
ds^2 = - \bar\sigma^2(\bar r) \bar w(\bar r) \, dt^2 + \frac{d
\bar r^2}{\bar w(\bar r)} + \bar r^2 \, d\Omega_2^2,
\end{equation}
and the GHS gauge \cite{Garfinkle:1990qj} $\sigma = 1$:
\begin{equation}
ds^2 = - w(r) \, dt^2 + \frac{dr^2}{w(r)} + \rho^2(r) \,
d\Omega_2^2.
\end{equation}
The coordinate transformation relating these gauges reads:
\begin{equation}
\bar r = \rho(r), \qquad \bar\sigma^2(\bar r) \bar w(\bar r) = w(r),
\qquad \frac1{\bar\sigma(\bar r)} = \frac{d \rho(r)}{dr}.
\end{equation}
It becomes singular at the turning points of the function $\rho(r)$
where the derivative $\rho'(r)=0$, so solutions containing such
turning points can not be described globally in the Schwarzschild
gauge. We will see later on that the presence of turning points is
typical for the DGB system, so  the GHS gauge is preferable.

\subsection{Dilaton black holes and the GB term}
In the theory without the GB term, $\alpha = 0$, an electrically
charged asymptotically flat black hole solution for an arbitrary
dilaton coupling $a$ reads (in the gauge $\sigma = 1$)
\cite{Garfinkle:1990qj}:
\begin{equation}\label{ghsa}
w(r) = \left( 1 - \frac{r_+}{r} \right) \left( 1 - \frac{r_-}{r}
\right)^{\frac{1-a^2}{1+a^2}}, \qquad \rho(r) = r \left( 1 -
\frac{r_-}{r} \right)^{\frac{a^2}{1+a^2}}, \qquad {\rm e}^{2 a \phi}
= {\rm e}^{2 a \phi_\infty} \left( 1 - \frac{r_-}{r} \right)^{-
\frac{2 a^2}{1+a^2}}.
\end{equation}
The mass and the electric charge are given by
\begin{equation}
M = \frac{r_+}2 + \frac{1-a^2}{1+a^2} \, \frac{r_-}2, \qquad q^2 =
{\rm e}^{2 a \phi_\infty} \frac{r_+ r_-}{1+a^2}.
\end{equation}
\par
For $a=0$ this reduces to the Reissner-Nordstr\"om solution, which
in the extremal limit $r_+ = r_- = r_H$,
\begin{equation} \label{ReNo}
ds^2 = - \left(1 - \frac{r_H}{r} \right)^2 dt^2 + \frac{dr^2}{
\left( 1 - \frac{r_H}{r} \right)^2} + r^2 \, d\Omega_2^2,
\end{equation}
has a degenerate event horizon $AdS_2\times S^2$. Note that for
$a=0$ the GB term decouples from the system, so this solution
remains true in the full theory with $\alpha \neq 0$.
\par
For $a \neq 0$,   the extremal limit $r_+ = r_- = r_H$  is:
\begin{equation}\label{edblh}
ds^2 = - \left( 1 - \frac{r_H}{r} \right)^{\frac2{1+a^2}} dt^2 +
\left( 1 - \frac{r_H}{r} \right)^{-{\frac2{1+a^2}}} \, dr^2 + r^2
\left( 1 - \frac{r_H}{r} \right)^{\frac{2 a^2}{1+a^2}} d\Omega_2^2.
\end{equation}
At $r = r_H$ the radius of the two-spheres shrinks, so we have a
point-like singularity. The Ricci scalar in the vicinity of this
point diverges near the horizon $r=r_H$:
\begin{equation}
R = \frac{2 a^2 r_H^2}{(1+a^2)^2 r^4} \left( 1 - \frac{r_H}{r}
\right)^{-\frac{2a^2}{1+a^2}},
\end{equation}
as well as the dilaton  exponential for an electric solution:
\begin{equation}
{\rm e}^{2 a \phi} = {\rm e}^{2 a \phi_\infty} \left( 1 -
\frac{r_H}{r} \right)^{- \frac{2a^2}{1+a^2}}.
\end{equation}
\par
Substituting the general dilatonic black hole solution (\ref{ghsa})
to the GB term we obtain the following value at the horizon $r=r_+$
\begin{equation}
\mathrm{e}^{2 a \phi}{\cal L}_{GB}|_{r=r_+} \sim (r_+ - r_-)^{-
\frac{2(4 a^2 + 1)}{a^2 + 1}}.
\end{equation}
This expression diverges in the extremal limit $r_+ \to r_-$. Thus,
it is not possible to treat the GB term perturbatively expanding in
$\alpha$ in the vicinity of the  extreme dilaton black hole. In
other words, one can expect that the GB term will substantially
modify the dilaton black hole solution in the extremal limit.  \par
Summarizing the above information, we see that the DGB gravity {\em
admits} the black hole solution with the degenerate event horizon of
the $AdS_2\times S^2$ type for $a=0$ (the Reissner-Nordstr\"om
solution (\ref{ReNo}), and {\em does not} admit the extremal dilaton
black hole (\ref{edblh}) for $a\neq 0$ even as an approximation for
small $\alpha$. So, the intriguing question arises, whether the
branch of {\em degenerate} black holes exists in the DGB gravity
which starts at $a=0$ in the parameter space and continues to
non-zero $a$. In the next section we analyze this possibility in
detail both analytically and numerically.

\section{DGB black holes with $AdS_2\times S^2$ horizon}
We are looking for asymptotically flat solutions in the DGB theory
for which the metric function $w(r)$ has double zero at some point
$r=r_H$ and does not have singularities for $r>r_H$. To attack this
problem numerically, one has to prove first that such solutions
exist  locally in the vicinity of the horizon $r=r_H$. We will show
that this is true, provided some restriction on the parameters is
satisfied.

\subsection{Near horizon expansion}

Assuming the GHS gauge $\sigma=1$,  consider the series expansions
around some point $r=r_H$ (supposed to be a  horizon)  in powers of
$x=r-r_H$:
\begin{equation}
w(r) = \sum_{k=1}^\infty w_k x^k, \qquad \rho(r) = \sum_{k=0}^\infty
\rho_k x^k, \qquad P(r) := {\rm e}^{2 a \phi(r)} = \sum_{k=0}^\infty
P_k x^k.
\end{equation}
The function $w$ starts with the linear term (vanishing of $w_0$
means that $r=r_H$ is a horizon), two other functions have general
Taylor's expansions.
\par
Substituting these expansions into the  equations of motion
(\ref{Eqf1}-\ref{Eqphi}) we find  local solutions of two types. The
first type solution has $w_1\neq 0$, i.e. the function $w$ has
simple zero at $r=r_H$. This corresponds to a non-degenerate horizon
of the Schwarzschild type. Such local solutions  and their numerical
continuation to infinity were considered in some particular cases in
refs. \cite{Kanti:1995vq, Torii:1996yi, Alexeev:1996vs,
Alexeev:1997ua}. Here we give more general expansion valid for both
electric and magnetic charges present (in the gauge $\sigma=1$):
\begin{eqnarray} \label{locnodeg}
w(r) &=& \frac{\Gamma}{\rho_0^2 P_1} \, x + \frac{2 a^2 \left[
\alpha (q_e^4 + P_0^2 q_m^2) + (4 \Gamma P_0 \alpha^2 + \rho_0^4)
q_e^2 + P_0^2 ( 4 \Gamma P_0 \alpha^2 - \rho_0^4 + 4 \alpha q_e^2)
q_m^2 \right] + \Gamma \rho_0^6}{12 \rho_0^6 a^2 P_0^2 \alpha} \,
x^2 + O(x^3),
\nonumber\\
\rho(r) &=& \rho_0 + \frac{(\rho_0^2 P_0 - q_e^2 - P_0^2 q_m^2 - 2
\alpha P_0 \Gamma) P_1}{\rho_0 P_0 \Gamma} \, x + O(x^2),
\nonumber\\
P(r) &=& P_0 + P_1 \, x + O(x^2),
\end{eqnarray}
where $\rho_0, P_0 $ and $P_1$ are free parameters and $\Gamma$
satisfies the equation
\begin{eqnarray}
& & 48 \alpha^3 a^2 P_0^2 \, \Gamma^2 + \left\{ \rho_0^6 - 16 P_0
\alpha^2 a^2 \left[ 3 \rho_0^2 P_0 - 2 (q_e^2 + P_0^2 q_m^2) \right]
\right\} \, \Gamma
\\
&+& 2 a^2 \left[ 2 \alpha (q_e^4 + P_0^4 q_m^4) + \rho_0^2 (\rho_0^2
- 12 P_0 \alpha) q_e^2 - \rho_0^2 P_0^2 (\rho_0^2 + 12 P_0 \alpha)
q_m^2 + 4 \alpha P_0^2 q_e^2 q_m^2 + 6 P_0^2 \alpha \rho_0^4 \right]
= 0. \nonumber
\end{eqnarray}
This quadratic equation has two roots $\Gamma= \Gamma_\pm $
depending on parameters $q_e, q_m, \rho_0, P_0, P_1$. The above
local solutions exists for such values of parameters for which
$\Gamma\neq 0$.
\par
The second class of local solutions has $w_1=0$. Vanishing of $w_1$
means that the horizon is degenerate.  Such an expansion contains
only one free parameter  with fixed charges. This family is
disconnected from the family (\ref{locnodeg}) and it was not noticed
so far:
\begin{eqnarray} \label{locdeg}
w(r) &=& \frac{ x^2}{\rho_0^2} - \frac{P_1}{6 \alpha a^2
  \rho_0^4} \left[ 3 (a^2-1) q_m^4 + 6 \alpha   (3a^2 - 2) q_m^2 + 4
  \alpha^2(5a^2 - 3)  \right]x^3 + O(x^4),
\nonumber \\
\rho(r) &=& \rho_0 + \frac{P_1}{4 \alpha a^2   \rho_0} \left[ (a^2 -
1) q_m^4 + 2 \alpha  (3 a^2 - 2) q_m^2 + 4\alpha^2  (a^2 - 1)
\right] x + O(x^2),
\nonumber \\
P(r) &=& \frac{\rho_0^2}{2 (2 \alpha + q_m^2)} + P_1 x + O(x^2).
\end{eqnarray}
Here $q_m$ is the magnetic charge, $P_1$ is a free parameter, and
$\rho_0$ is the physical radius of the horizon depending on charges
as follows
\begin{equation}
\rho_0^2 = \frac{ 2 q_e (2\alpha + q_m^2) }{\sqrt{4 \alpha  + q_m^2}
}.
\end{equation}
Note that the dilaton coupling constant enters these expansion only
through $a^2$, so the space of solutions is symmetric under $a\to
-a,\,\phi\to -\phi$. In what follows we will assume $a\geq 0$.

The values of the integrals of motion corresponding to
(\ref{locdeg}) are
\begin{eqnarray}
H &=& \frac1{2\rho^2} \left[ q_m^2 P_0 + q_e^2 P_0^{-1} - \rho^2
\right],
\qquad P_0 = \frac{\rho_0^2}{2 (2 \alpha + q_m^2)},
\\
J_\delta &=& 2 \, q_e \, f_0,
\end{eqnarray}
where $f_0$ is the value of the electrostatic potential on the
horizon.
\par
From our previous analysis of the Einstein-Maxwell-Dilaton (EMD)
black holes and the above form of the local solution one can draw
the following conclusions:
\begin{enumerate}
\item Black holes with  $AdS_2\times S^2$ event horizon do  not
exist in the $a\neq 0$ EMD theory without curvature corrections
($\alpha=0$).
\item
Such solutions do not exist in absence of the electric charge, while
the presence of the magnetic charge is optional. S-duality is thus
broken as expected.
\item
This local solution is not generic (the number of free parameters is
less that the degree of the system of differential
equations).
\end{enumerate}
\par
For simplicity, in this paper we will focus on  purely electric
black holes. In this case the expansions simplify and we can
give some further terms:
\begin{eqnarray} \label{expanElec}
w(r) &=& \frac1{\rho_0^2} \left[ x^2 - \frac{2 (5 a^2 - 3)}{3}
\left( \frac{\alpha P_1}{ a^2\rho_0^2} \right) x^3 + \frac{173 a^6 -
269 a^4 + 99 a^2 - 27}{3 (5 a^2 - 3)  } \left( \frac{\alpha P_1}{
a^2\rho_0^2} \right)^2 x^4 \right] + O(x^5),
\nonumber \\
\rho(r) &=& \rho_0 \left[ 1 +  ({a^2 - 1}) \left( \frac{\alpha
P_1}{{a^2}\rho_0^2} \right) x - \frac{2 a^2 (a^4 - 6)}{ (5 a^2 - 3)}
\left( \frac{\alpha P_1}{{a^2}\rho_0^2} \right)^2 x^2 \right] +
O(x^3),
\nonumber \\
\alpha P(r) &=& \rho_0^2 \left[ \frac14  + a^2 \left(
\frac{\alpha P_1}{a^2 \rho_0^2} \right) x + \frac{a^2(a^4 - 5 a^2 -
3)}{ (5a^2 - 3)} \left( \frac{\alpha P_1}{ a^2 \rho_0^2} \right)^2
x^2 \right] + O(x^3).
\end{eqnarray}
The electric charge  is related to the horizon radius as
\begin{equation} \label{defqe}
q_e = \frac{\rho_0^2}{{2 \sqrt \alpha} }.
\end{equation}
We can arrange higher order terms in such a way, that $a^2$ enters
in denominator only in powers of the combination $P_1/a^2$. This
facilitates taking the limit $a\to 0$. We know that in this limit
there exists an exact solution which has the near-horizon expansion
of the type (\ref{expanElec}), namely the extremal
Reissner-Nordstr\"om solution. Thus we expect that for the
asymptotically flat solutions
\begin{equation}
P_1 \to 0, \quad {\mbox{as}} \quad a \to 0.
\end{equation}
Numerical calculations  show that this is indeed the case (see Sec.
C).
\par
Another subtlety is related to the limit of GB decoupling.
Obviously, our expansion fails in this limit: for a finite charge
$q_e$ one has $\rho_0^2 = 2 q_e \sqrt \alpha \to 0$ and,
consequently, the expansion coefficients will diverge. The reason is
that our expansion for $w$ is incompatible with that for the dilaton
black hole of the Einstein-Maxwell-Dilaton theory. This reflects
again the absence of the black hole solutions with the $AdS_2\times
S^2$ horizon  in the theory without curvature corrections.
Substituting $q_e$ defined by (\ref{defqe}) into the equations of
motion (\ref{Eqf1}-\ref{Eqphi}), one can see that the GB coupling
parameter $\alpha$ enters  always in the combination $\alpha {\rm
e}^{2 a \phi(r)}$. Thus, shifting the dilaton is equivalent to
rescaling the GB term (this was in fact clear already from the
action (\ref{reducedS})). Note that in the case $a^2 = 1$ the linear
term in the expansion of $\rho$ vanishes, $\rho_1 = 0$, implying
that there is no regular transformation to the gauge $\rho = \bar
r$.
\par
Therefore, a purely electric local solution with a fixed value of
charge $q_e$ contains one free parameter:  the dilaton derivative
$P_1$. An important issue is to determine the correct sign of $P_1$.
To be able to interpret the region $r>r_H$ as an exterior of the
black hole, one has to ensure positiveness of the derivative $\rho'$
at the horizon. From the above expansion one finds
\begin{equation}
\rho'|_{x=0} = \frac{(a^2 - 1) \alpha P_1}{a^2 \rho_0} > 0.
\end{equation}
Thus, we should take positive $P_1$ for $a^2 > 1$ and negative $P_1$
for $a^2 < 1$. It is convenient to introduce the sign parameter
which ensures this:
\begin{equation}
\varsigma = \frac{P_1}{|P_1|} = \frac{a^2-1}{|a^2-1|}.
\end{equation}
\par
Now define the following combination of $\rho_0$ and  $P_1$:
\begin{equation}
b = \frac{\alpha |P_1|}{a^2\rho_0^2}.
\end{equation}
It is easy to see, that free parameters enter the series expansions
near the horizon only through this quantity. Consider now the
transformations of the expansion parameters under symmetries of the
solution space (\ref{trr},\ref{symsol}). Since $P_1$ is the first
derivative of the dilaton exponent, one finds that under
$\delta$-transformation
\begin{equation}
P_1 \to P_1 \mathrm{e}^\delta, \quad b \to b \mathrm{e}^{-\delta},
\end{equation}
so the quantity $bx$ remains invariant. Thus the full set of local
solutions can be generated from one particular solution with
$\rho_0=1,\, P_1=1$, which we will call the normalized local
solution,  by the symmetry transformations with $\delta = - \ln
\rho_0$ and $\mu = 2 \ln (b \rho_0)$ from (\ref{symsol}), i.e. by
the group element $G(2 \ln (b\rho_0), - \ln \rho_0)$.  The
normalized local solution does not contain free parameters at all:
\begin{eqnarray}
w(r) &=& x^2 - \varsigma \frac{2 (5a^2-3)}{3 }x^3 + \frac{(173
a^6 - 269 a^4 + 99 a^2 - 27)}{3 (5 a^2 - 3)} x^4 + O(x^4),
\label{fwr} \\
\rho(r) &=& 1 + \varsigma  (a^2 - 1)  x - \frac{2a^2(a^4- 6)}{
(5 a^2 - 3)}x^2 + O(x^3), \label{fro}
\\
\alpha P(r) &=& \frac14 + \varsigma (r - r_H) + \frac{(a^4 - 5 a^2 -
3)}{ (5a^2 - 3)}x^2 + O(x^3). \label{fhe}
\end{eqnarray}
Note  the presence of the sign function  $\varsigma$ in the odd
power terms. The electric charge corresponding to the normalized
local solution is $q_e = 1/(2\sqrt{\alpha})$.

\subsection{Asymptotic flatness}
We are looking for asymptotically flat global solutions which
satisfy the conditions $w \to 1,\; \rho/r \to 1, \;\phi \to {\rm
const}$ as $r \to \infty$. The subleading terms should be expandable
in the power series of $1/r$. The local solution with
these properties turns out to be three-parametric, depending on the
ADM mass $M$, the dilaton charge $D$, and the asymptotic value of
the dilaton $\phi_\infty$:
\begin{eqnarray}\label{asym}
w(r) &=& 1 - \frac{2M}r + \frac{\alpha Q_e^2}{r^2} + O(r^{-3}),
\nonumber\\
\rho(r) &=& r - \frac{D^2}{2 r} - \frac{D(2 M D - \alpha a Q_e^2)}{3
r^2} + O(r^{-3}),
\\
\phi(r) &=& \phi_{\infty} + \frac{D}r + \frac{2 D M - \alpha a
Q_e^2}{2 r^2} + O(r^{-3}), \nonumber
\end{eqnarray}
where \vspace{-.5cm}
\begin{equation}
Q_e = q_e \mathrm{e}^{-a\phi_\infty}.
\end{equation}
The dilaton charge  can be also read  from the asymptotic expansion
of the dilaton exponential:
\begin{equation}
\mathrm{e}^{2 a (\phi - \phi_\infty)} = 1 + \frac{2 a D}r + \frac{2
a D (a D + M) - \alpha  a^2 Q_e^2}{r^2} + O(r^{-3}).
\end{equation}
The asymptotic values of two integrals of motion are:
\begin{eqnarray}
H &=& \frac12 \left( w_\infty \rho'^2_\infty - 1 \right),
\label{J1a} \\
J_\delta &=& 2 \, q_e \, f_{\infty} - M - \frac{D}a.
\end{eqnarray}
\par
Behavior of the global solution which starts with the normalized
local solution (\ref{fwr},\ref{fro},\ref{fhe}) at the horizon
depends only on the dilaton coupling constant $a$. Its existence for
all $a$ is not guaranteed {\em a priori}. But, for some sufficiently
small values of  $a$, we find numerically that all three functions
vary smoothly with increasing $x$, so that $w$ and the derivative
$\rho'$ stabilize at infinity on some constant values $w_\infty \neq
1,\, \rho'_\infty \neq 1$. Then, using the symmetries
(\ref{trr},\ref{symsol}) of the solution space, one can rescale the
whole solution to achieve the desired unit values for these
parameters. More precisely, the relevant subgroup of rescalings is
two-parametric: $G(\mu,\delta)$. As we have argued, two parameters
$\mu, \delta$ effectively replace the parameters $\rho_0, P_1$ of
the (non-normalized) local solution (\ref{expanElec}). So one could
expect that rescaling of the solution so that $w_\infty = 1,\,
\rho'_\infty = 1$ would fix both  quantities $\rho_0, P_1$ on the
horizon. But from the Hamiltonian equation $H=0$ with $H$ given by
the Eq. (\ref{J1a}) it is easy to see that one must have $w_\infty
\rho'^2_\infty = 1$ for any solution such that $w\to
w_\infty,\;\rho'\to\rho'_\infty$ asymptotically. Therefore it is
enough to perform {\em one} but not {\em two} independent rescalings
in order to get $w_\infty = 1,\, \rho'_\infty = 1$. Indeed, under
$G(\mu,\delta)$
\begin{equation}
w \to w \mathrm{e}^\mu, \quad \rho_0 \to \rho_0 \mathrm{e}^\delta,
\quad P_1 \to P_1 \mathrm{e}^{\delta - \mu/2}, \quad w \rho'^2 \to w
\rho'^2.
\end{equation}
Since the choice of $\mu, \delta$ is equivalent to the choice of
$\rho_0,\, P_1$, an invariance of the product $w \rho'^2$ under
$G(\mu, \delta)$ means that the solution starting on the horizon
with {\em any} $\rho_0, P_1$ will reach at infinity the values
$w_\infty,\, \rho'_\infty$ satisfying $w_\infty {\rho'}^2_\infty =
1$. Therefore, taking $\mu = - \ln w_\infty$, we will achieve
simultaneously $w_\infty = 1$ and $\rho'_\infty = 1$. This means
that asymptotically flat solutions still form a one-parameter
family, a parameter being the electric charge $q_e$.

\subsection{Numerical analysis}
Since we know that the desired global solution exists for $a=0$, we
start with the local solution at the horizon with small $a$ and look
for numerical solutions which fit the asymptotic expansions
(\ref{asym}). For sufficiently small $a$ global solutions exist
indeed, and, as we have explained, two basic conditions at infinity
$w = 1,\, \rho'=1$ fix only one of the two parameters $\rho_0$ and
$P_1$ at the horizon. It will be convenient to leave $\rho_0$
(defining the electric charge) arbitrary, and to fix $P_1$. We will
also choose  the value of the GB coupling $\alpha = 1.$ Then the
black hole mass can be found numerically from the asymptotic
expansions (\ref{asym}) together with the dilaton charge and the
asymptotic value of the dilaton.
\par
Typical coordinate dependence of the metric functions and the
dilaton exponent are shown in Fig. \ref{aff} for some values of the
dilaton coupling $a$. Solutions exist for
\begin{equation}
0 \leq a < a_{\rm cr}, \quad a_{\rm cr} \simeq 0.488219703.
\end{equation}
\psfrag{r1}{\large$x$} \psfrag{w}
{\large$w(x)$}\psfrag{rho'}{\large$\rho'(x)$}\psfrag{F}{\large$P(x)$}
\psfrag{a1}{\scs $a=0.1$}\psfrag{a4}{\scs
$a=0.4$}\psfrag{a45}{\scs$a=0.45$}
\begin{figure}[h]
\vbox to 8cm{ \vfil \hbox to 18cm{\hspace{-.5cm} \hfil
\includegraphics[height=7cm]{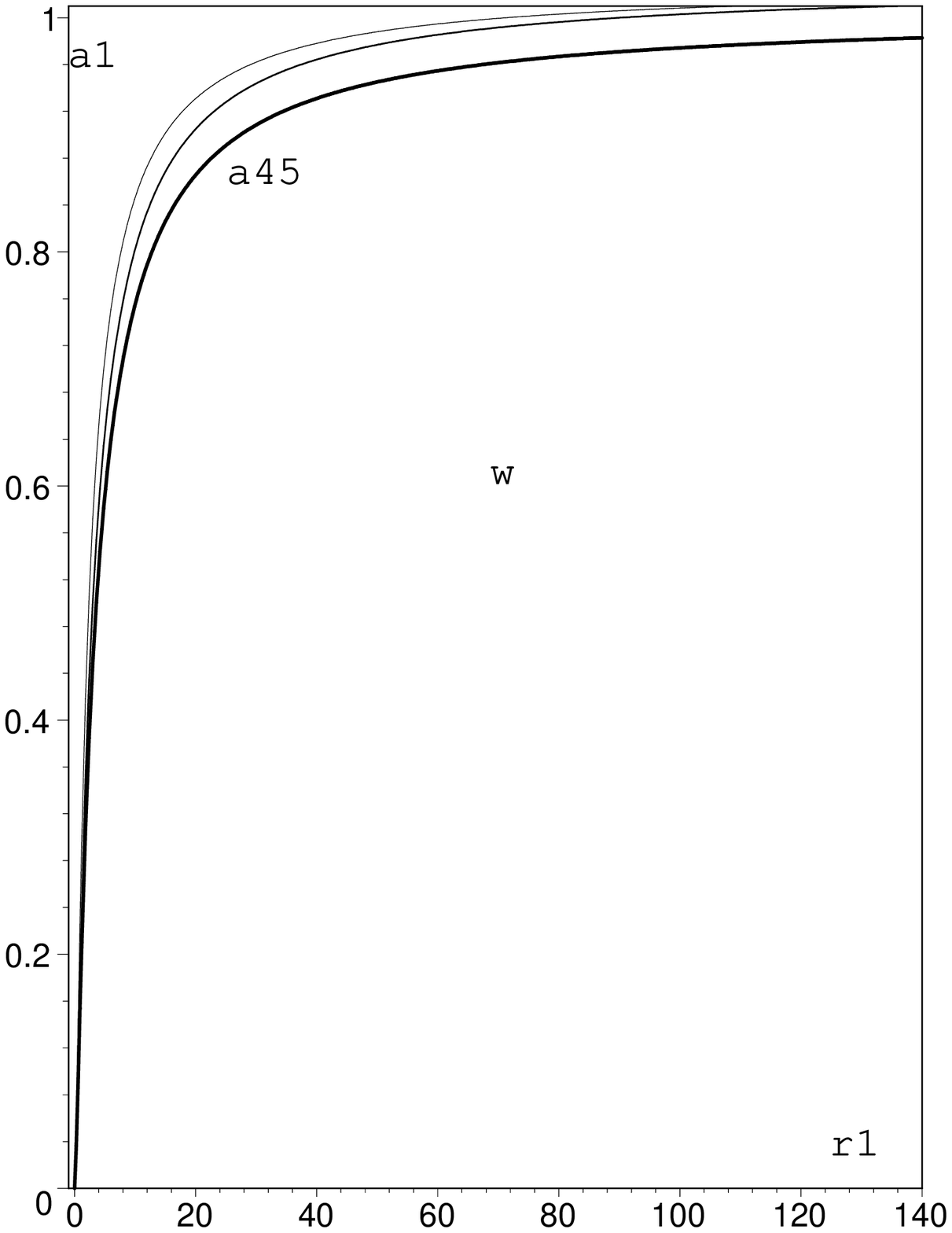}
\hfil
\includegraphics[height=7cm]{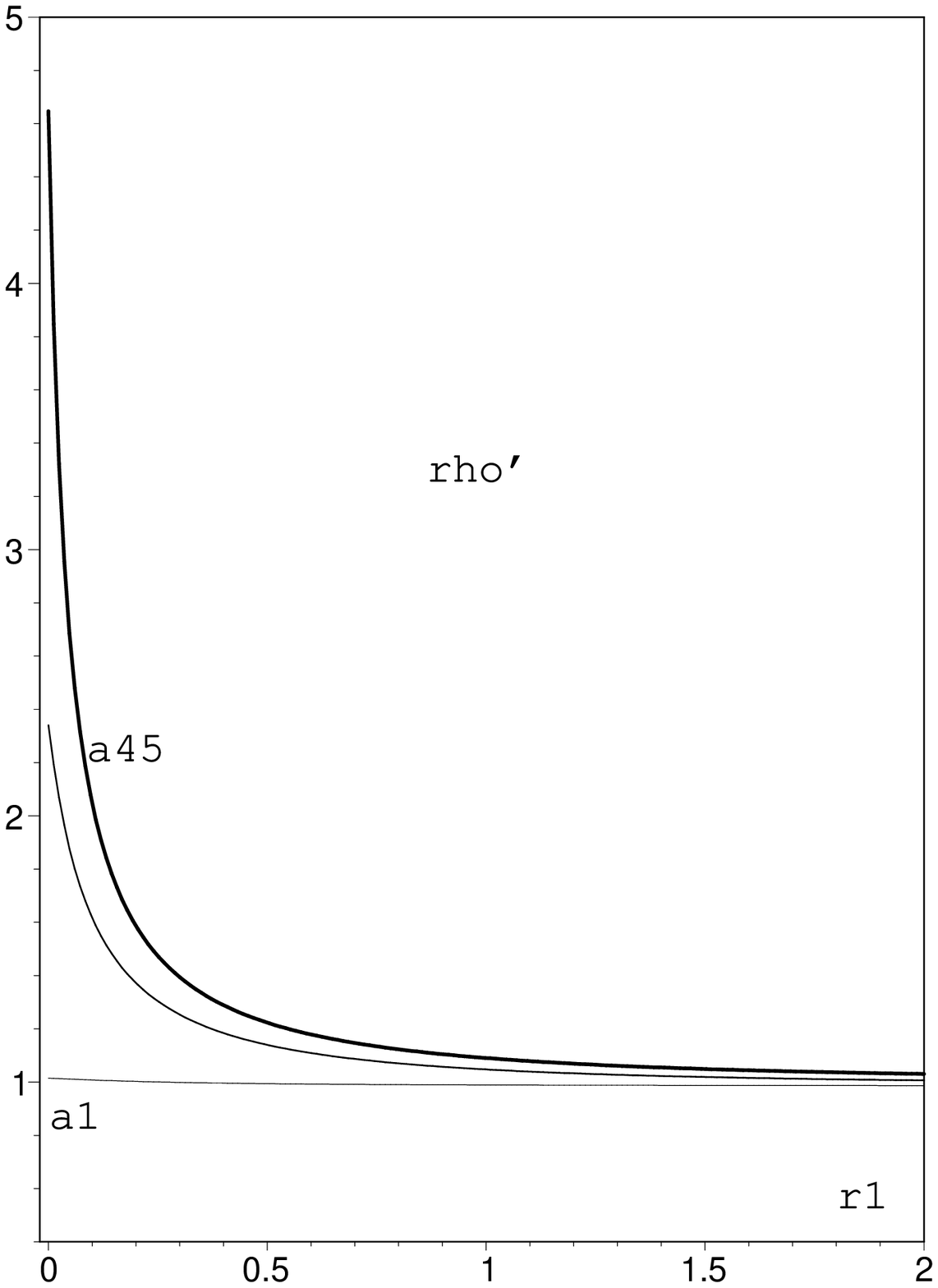}
\hfil
\includegraphics[height=7cm]{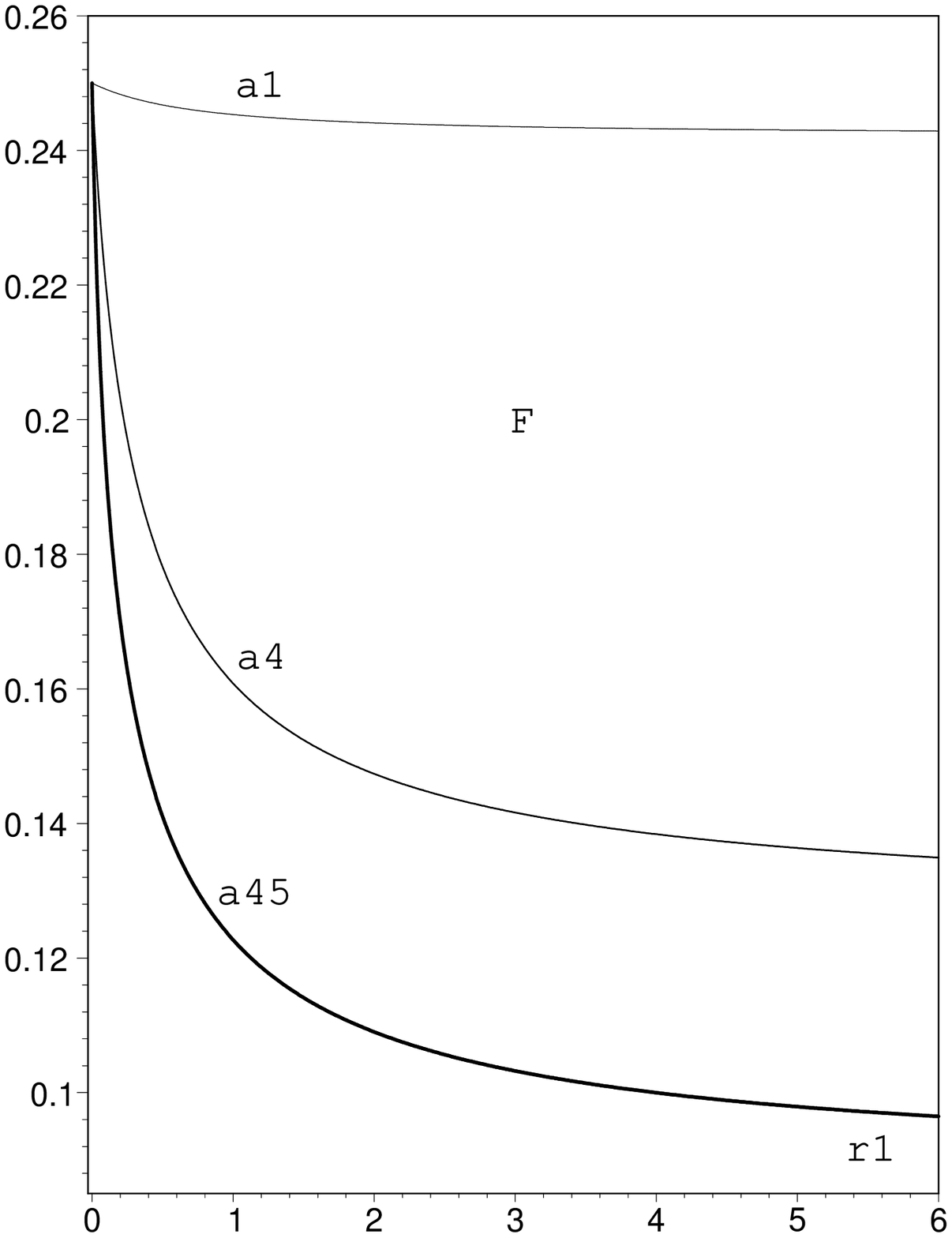}
\hfil} \vfil } \caption{The functions $w(x), \rho'(x),
P(x),\;x=r-r_H,$ for  $\rho_0=1$ and some values of the dilaton
coupling constant: $a=0.1$ ($P_1=-0.01$) ---  thin line, $a=0.4$
($P_1=-0.446$) --- normal line, $a=0.45$ ($P_1=-1.18$) --- thick
line.} \label{aff}
\end{figure}
Let us discuss in more detail behavior of solutions at the ends of
this interval. As expected, the parameter $b$ of the local solution
at the horizon tends to unity when $a\to 0$, as shown in Fig.
\ref{bf}. This means that the first Taylor coefficient in the
expansion of $\rho(x)$ becomes equal to unity (note that the sign
function $\varsigma=-1$ as $a\to 0$), while all higher coefficients
are zero. Therefore, assuming $\rho_0=r_H$, we find that $\rho=r$
globally. Similarly, all terms in the expansion of $P(x)$ vanish in
the limit $a\to 0$ except the constant $P_0$, so the dilaton
exponential tends to the constant value $P = \rho_0^2/4$.
Correspondingly,  we find
\begin{equation}
\lim_{a \to 0} \mathrm{e}^{2 a \phi_\infty} / q_e \to 1/2.
\end{equation}
For $w$  all Taylor's coefficients in (\ref{expanElec}) are non-zero
and the whole series exactly reproduces an expansion
\begin{equation}
w(r) = \left( 1 - \frac{\rho_0}{r} \right)^2 = z^2 - 2 z^3 + 3 z^4 +
O(z^5), \qquad z = (r - \rho_0)/\rho_0.
\end{equation}
Thus, our family of solutions begins with the extremal
Reissner-Nordstr\"om metric for zero dilaton coupling $a$.

\psfrag{b}{\hspace{1.3cm}\raisebox{-1cm}{$b(a)$}}
\psfrag{a}{\hspace{-0.1cm} $a$}
\psfrag{acr}{\hspace{-0.1cm}\fns $a_{cr}$}
\psfrag{eq1}{\hspace{-0.1cm}$\left|\frac{aM}{D}\right|$}
\psfrag{eq2}{\hspace{-0.6cm}$ \frac{\sqrt{(1+a^2)}M}{Q_{e}}$}
\psfrag{M2Q}{\large $\frac{M^2}{q_e}$} \psfrag{D2Q}{\large
$\frac{D^2}{q_e}$} \psfrag{FQ}{\hspace{-0.2cm}\large
$\frac{\e^{2a\phi_{\infty}}}{q_e}$}
\begin{figure}
\vbox to 7cm{ \vfil \hbox to 18cm{\hspace{-.5cm}
\hfil
\includegraphics[height=7cm]{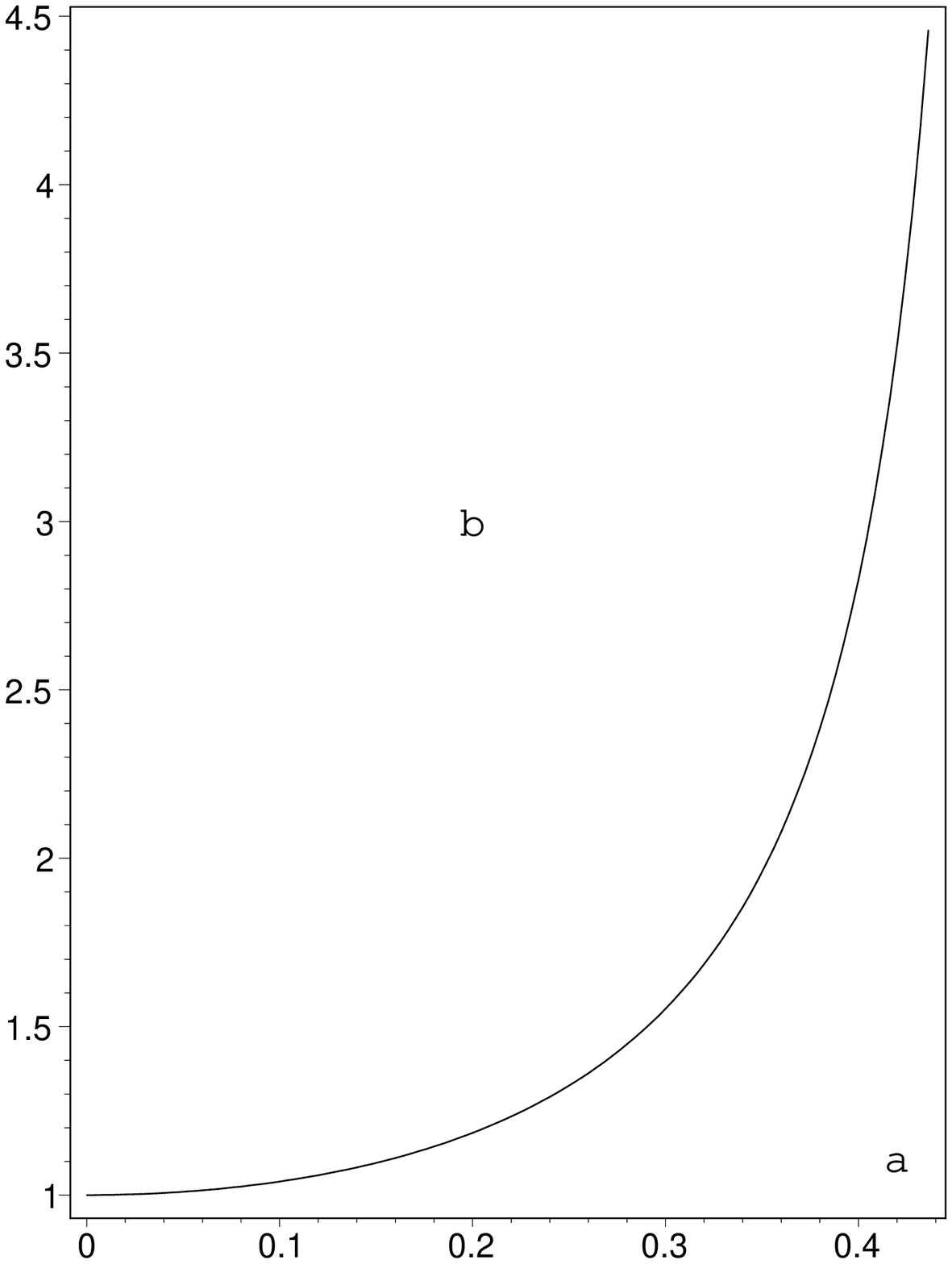}
\hfil
\includegraphics[height=7cm]{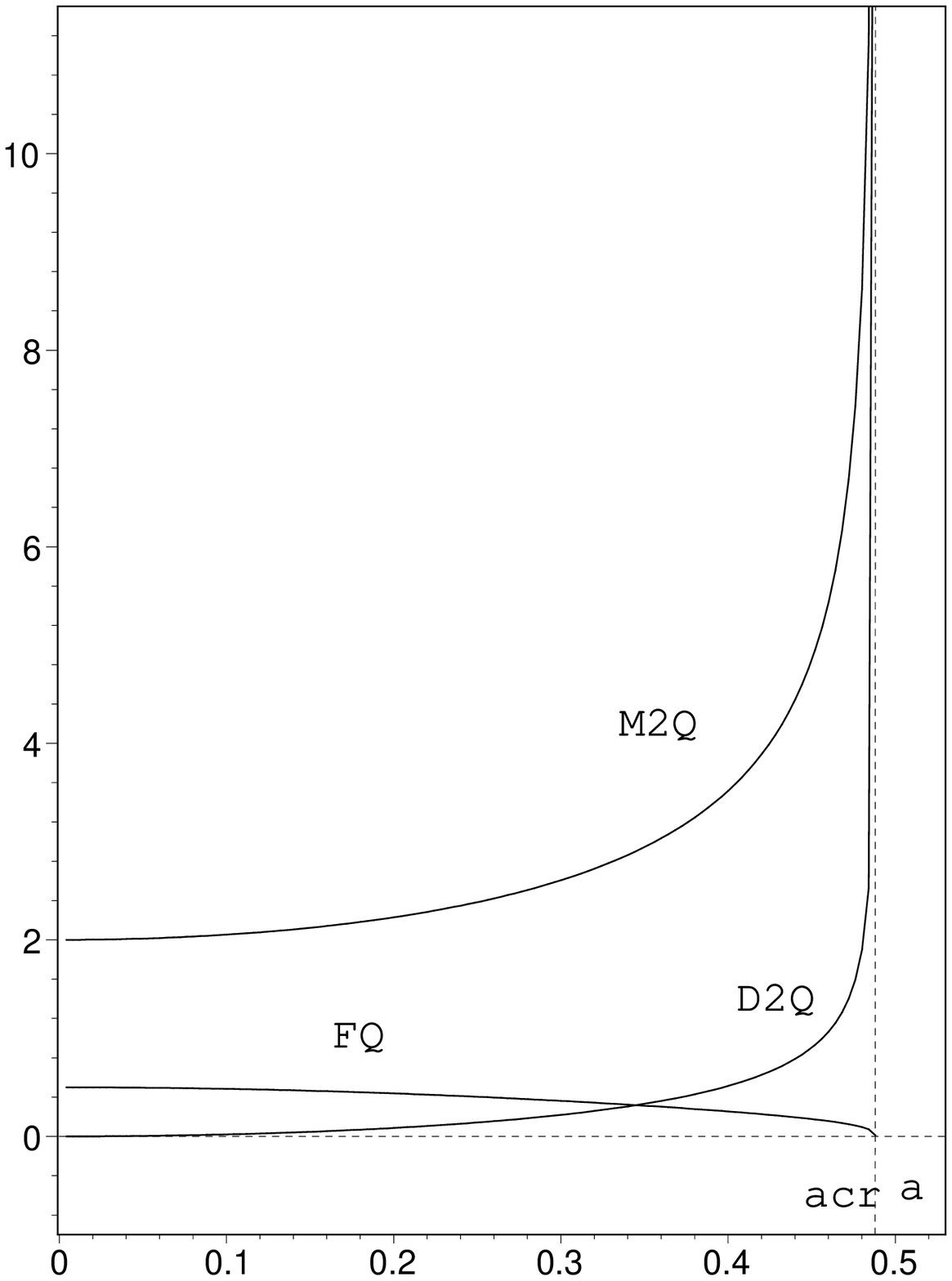}
\hfil
\includegraphics[height=7cm]{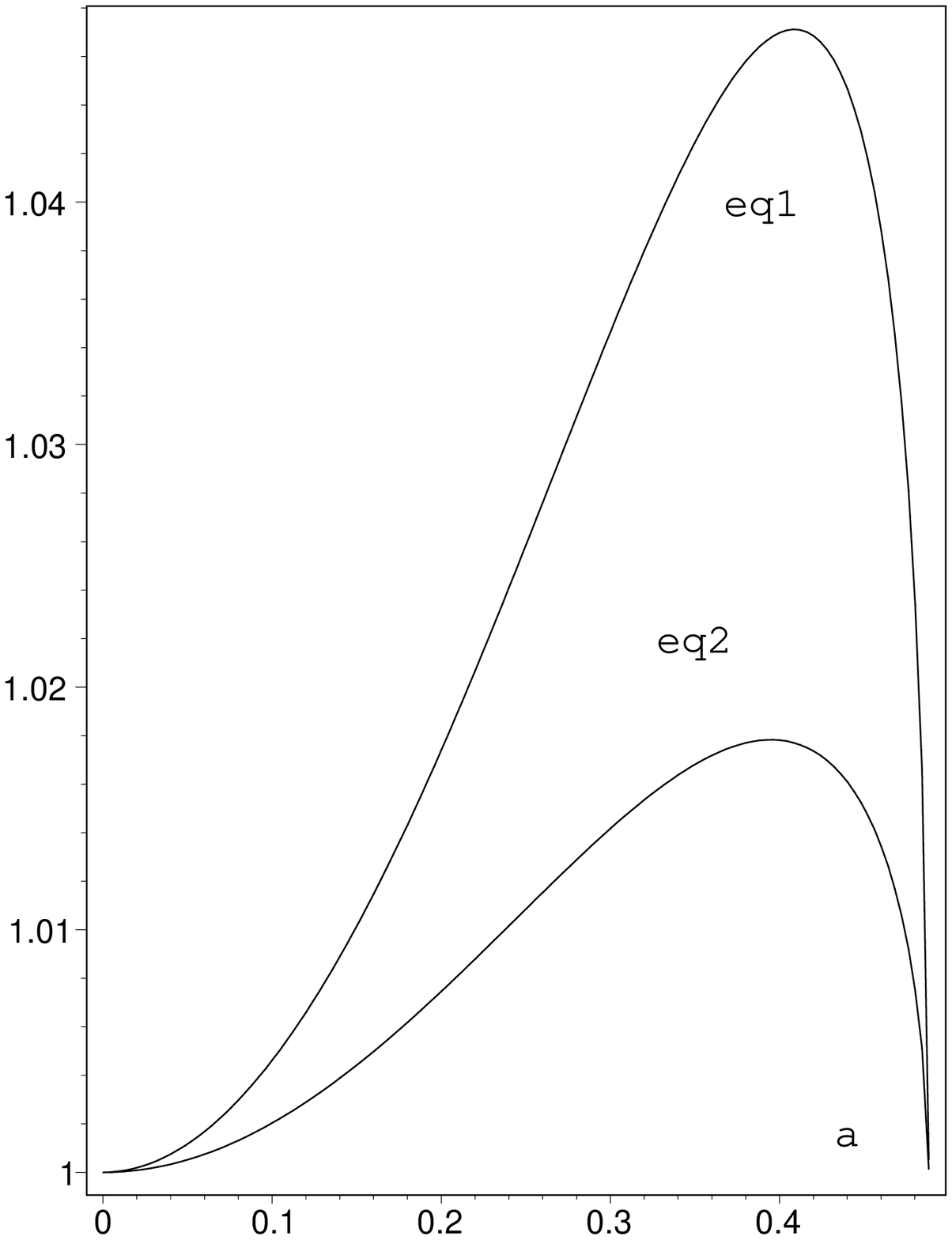}
\hfil}
\vfil }
\renewcommand{\thefigure}{2.1}
\caption{Dependence of the parameter $b$ on the dilaton coupling
constant $a$: $b$ tends to unity for $a\to 0$ ensuring  continuous
transition to the extremal Reissner-Nordstr\"om solution.}
\label{bf}
\renewcommand{\thefigure}{2.2}
\caption{Numerical curves $k_M(a)= M^2/q_e , \;k_D(a)= D^2/q_e
,\;k_\phi(a)=\e^{2a\phi_\infty}/q_e $ for $\rho_0=1$. The mass curve
starts with the Reissner-Nordstr\"om value for $a=0$ and
diverges as $a\to a_{\rm cr}$. The dilaton
charge increases from zero to infinity, while the dilaton
exponential monotonically varies from the value $1/2$ at $a=0$ to
zero for $a\to a_{\rm cr}$.} \label{Qdf}
\renewcommand{\thefigure}{2.3}\addtocounter{figure}{-2}
\caption{The quantities $\left|\frac{\ds  aM}{\ds D}\right|$ and
$\frac{\ds \sqrt{ 1+a^2 }M}{\ds Q_e}$ as functions of $a$: both tend to
unity at the ends of the allowed interval of $a$.}
\label{BPScf}
\end{figure}
\par
With fixed horizon radius $\rho_0$, the mass and the dilaton charge
of the black holes increase with the growing dilaton coupling
constant tending to infinity when $a$ approaches $a_{\rm cr}$. The
dilaton exponent, on the contrary, tends to zero in this limit.
Using the symmetry of the solution space under
$\delta$-transformation, one can generate the sequence of solutions
with different electric charges $q_e$ and correspondingly with
different masses, dilaton charges and $\phi_\infty$. Since variation
of the electric charge is essentially equivalent to variation of the
unique parameter $\rho_0$ in the horizon expansion, it is clear,
that using $\delta$-transformation  we will generate {\em all}
extremal solutions. Under this transformation the  mass and the
dilaton charge  scale as $\mathrm{e}^{\delta}$, while the electric
charge and the dilaton exponent $\mathrm{e}^{2a \phi_\infty}$ scale
as $\mathrm{e}^{2\delta}$. Therefore the ratios
\begin{equation}
k_M = \frac{M^2}{q_e}, \quad k_D = \frac{D^2}{q_e}, \quad k_\phi =
\frac{\mathrm{e}^{2a\phi_\infty} }{q_e}
\end{equation}
depend only on $a$. Their numerical graphs are presented on
Fig.\ref{Qdf}.
\par
As we already discussed, the metric for $a=0$ is known analytically.
For $a$ in the vicinity of $a_{\rm cr}$ the analytic solution is not
known, but one finds that the behavior of the mass, the dilaton
charge and the rescaled electric charge $Q_e$ exhibits some
similarities at both ends of the allowed interval of $a$.  Namely,
the following two ratios stabilize at unity for both $a \to 0$ and
$a \to a_{\rm cr}$ (Fig. \ref{BPScf}):
\begin{equation}
\left| \frac{a M}{D} \right| \to 1, \qquad \frac{\sqrt{ 1 + a^2
}M}{Q_e} \to 1
\end{equation}
This corresponds to fulfillment of the following condition
\begin{equation}\label{BPScond}
a^2 M^2 + D^2 = \frac{2 a^2}{ 1 + a^2 } \, Q_e^2,
\end{equation}
which is reminiscent of the BPS condition for electrically charged
black holes of the EMD theory. This feature is  similar to that in
another stringy generalization of EMD theory in which the Maxwell
action is replaced by the Born-Infeld action, but no GB term is
introduced \cite{Clement:2000ue}.

\psfrag{r1}{\large $x$}\psfrag{rho1}{\large $\rho$}
\psfrag{w(rho)}{\large$w(\rho)$}\psfrag{rho}{\large$\rho(x)$}
\psfrag{r0}{\scs $\rho_t$}\psfrag{rs}{\scs $\rho_s$}\psfrag{rf}{\scs
$\rho_f$}
\begin{figure}
\vbox to 7cm{ \vfil \hbox to 18cm{\hspace{-2cm} \hfil
\includegraphics[width=5cm]{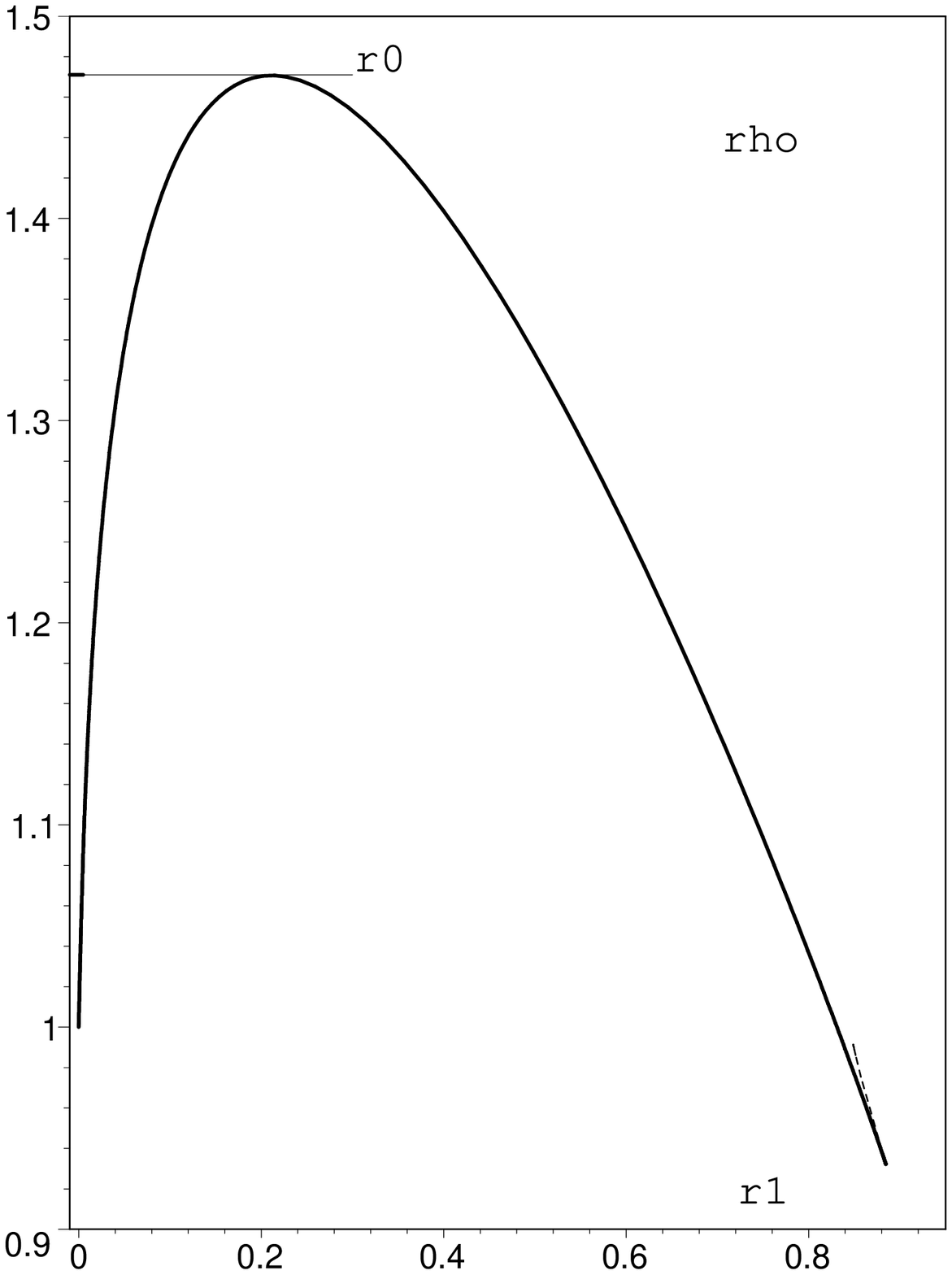}
\hfil
\raisebox{0.7cm}{\hspace{-7.8cm}\includegraphics[width=3.5cm]{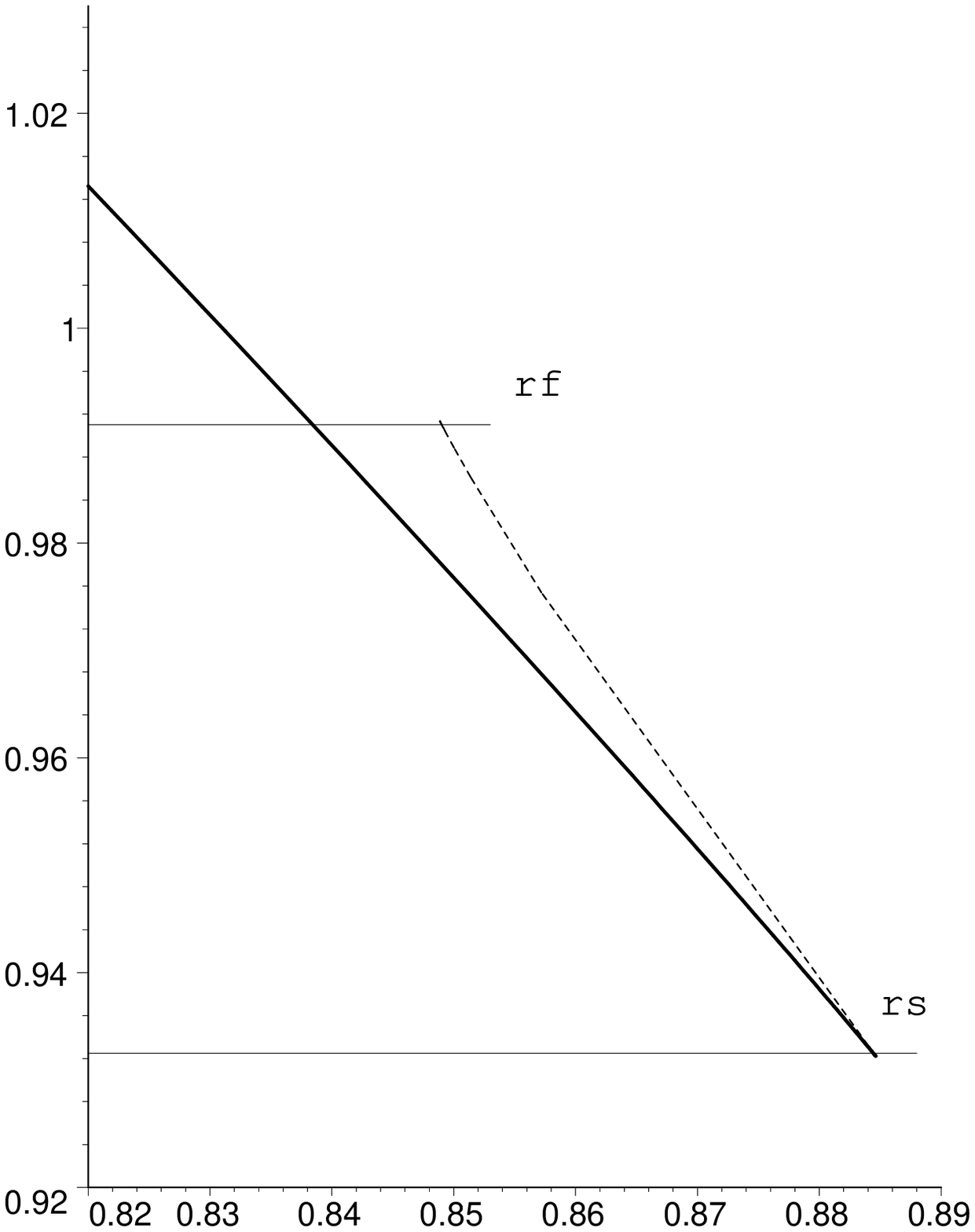}}
\hfil
\includegraphics[width=5cm]{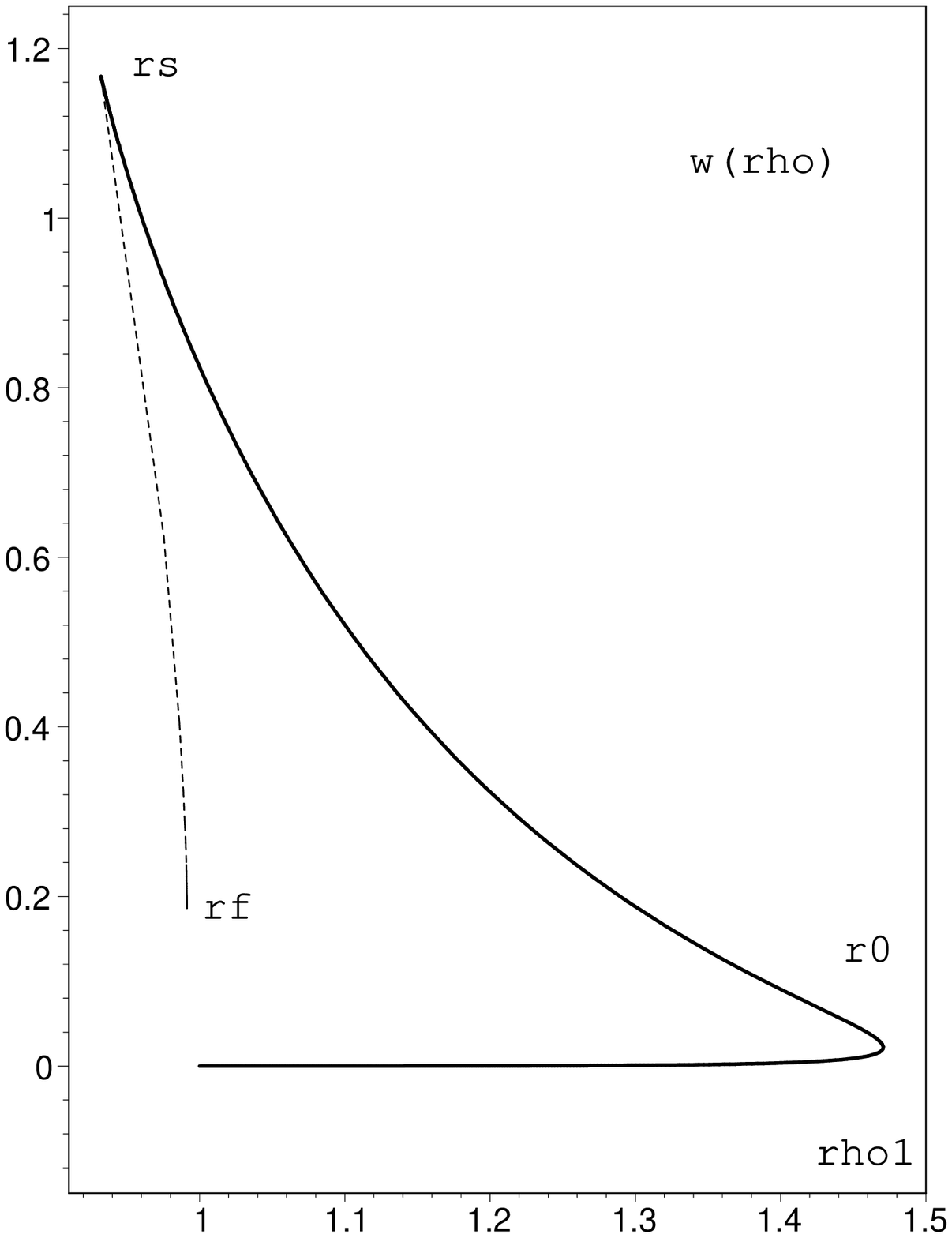}
\hfil} \vfil } \caption{The metric functions $\rho(x), w(\rho),\, (x=r-r_H)$
for the value of the dilaton coupling constant $a>a_{\rm cr}$. The function
$\rho$ has a turning point $r=r_t$ after which the naked singularity
is met at a finite affine distance ($r=r_s$). Dotted lines correspond  to
another solution branch which can be matched in the singularity. Numerical
curves are presented for $a=0.5$ and  $  \rho_0=1$, the
corresponding value of $P_1$ being $P_1=-7.746$.} \label{naff}
\end{figure}

For the values of  $a$ outside the allowed interval, solutions
starting with the $AdS_2\times S^2$ horizon are not asymptotically
flat, but singular. For them the metric function $\rho(r)$ has a
turning point at some finite radial coordinate $r=r_t$, such that
$\rho'(r_t)=0,\, \rho''(r_t)<0$ (Fig. \ref{naff}). This point is
regular, but at a finite proper distance from it one encounters the
singular turning point $r_s$, such that $\rho'(r_s)>0,\,
\rho''(r_s)=\infty$, where all variables have a square-root
singularity, being expandable in terms of $\sqrt{r-r_s}$:
\vspace{-.2cm}
\begin{eqnarray}
w &=& w_s + w_1 y + w_{3/2} \; y^{3/2} + O(y^2), \qquad y = r - r_s,
\nonumber\\
\rho &=& \rho_s + \rho_1 y + \rho_{3/2} \; y^{3/2} + O(y^2),
\\
P &=& P_s + P_1 y + P_{3/2} \; y^{3/2} + O(y^2) \nonumber.
\end{eqnarray}
This local expansion contains four free parameters $w_s,\, \rho_s,\,
\rho_1,\,P_s$, while other coefficients read:
\begin{eqnarray}
w_1 &=& \frac{4 a^2 P_s (\gamma \rho_s^2 P_s + q_e^2) - \Delta^2 w_s
\rho_s^4}{4 P_s^2 a^2 \rho_s^2 \left[ \rho_1 (6 \Delta w_s \rho_1 -
\rho_s) - 2 \Delta \right]}, \quad P_1 = \Delta, \quad P_{3/2} =
\frac{(\rho_s - 4 \Delta \rho_1 w_s) \rho_{3/2}}{2 \gamma},
\\
w_{3/2} &=& \frac{\left\{ 2 \Delta \rho_s^4 \left[ 2 \Delta \rho_1
w_s (\gamma + 4) - \rho_s (5 \gamma + 4) \right] - 16 a^2 P_s \rho_1
\gamma (\gamma \rho_1 P_s \rho_s^2 + q_e^2) + \rho_s^6 \rho_1
\right\} w_s \rho_{3/2}}{8 \rho_s^2 a^2 P_s^2 \gamma^2 (2 \Delta (3
\gamma + 2) - \rho_s \rho_1)},
\end{eqnarray}
where $\gamma = \rho_1^2 w_s - 1$ and $\Delta$ satisfies the
equation:
\begin{eqnarray}
&& \Delta^3 \left\{ 8 w_s \rho_s^4 \left[ w_s \rho_1^2 (15 \gamma +
9) - 1 \right] \right\} - \Delta^2 24 w_s \rho_s^5 \rho_1 (3 \gamma
+ 2) + \rho_s \rho_1 \left[ q_e^2 32 a^2 P_s \gamma + \rho_s^2 (48
a^2 P_s^2 \gamma^2 - \rho_s^4) \right]
\\
&+& \Delta \left\{ 2 \rho_s^2 \left[ \rho_s^4 (\gamma + 6) + 96
\rho_1^2 w_s a^2 P_s^2 \gamma^2 \right] - q_e^2 32 a^2 P_s (3 \gamma
+ 4) \gamma \right\} = 0. \nonumber
\end{eqnarray}
An expression for $\rho_{3/2}$ is too big and is not given here.
Since the second derivatives are divergent at $y=0$, the Riemann
tensor diverges as well. The divergency is localized on a sphere of
finite radius, and it is rather mild: Ricci and Kretchmann scalars
behave as
\begin{equation}
R \simeq - \frac{3 (4 \rho_{3/2} w_s + w_{3/2} \rho_s)}{4 \rho_s }\,
\frac1{\sqrt{y}}, \qquad R_{\alpha\beta\gamma\delta}
R^{\alpha\beta\gamma\delta} \simeq \frac{9 (8 \rho_{3/2}^2 w_s^2 +
\rho_s^2 w_{3/2}^2)}{16 \rho_s^2}\,\frac1{y}.
\end{equation}
The radial coordinate stops at $r=r_s$, but using an appropriate
desingularization of the system (see Appendix), one can glue another
patch of radial coordinate $r'\in(r_s,\, r_f)$ at this point,
extending the manifold through the singularity. This extension is
shown by dotted lines in Figs. \ref{naff}. It terminates at the
final singularity $r_f$.  This situation is very similar to that
described in ref. \cite{Alexeev:1997ua} for an interior region of
the non-extremal DGB black hole.

\subsection{Thermodynamics}
The temperature of the extremal DGB black hole  is zero,
as for the extremal solution without the GB term:
\begin{equation}
T = \frac1{2 \pi} \left. \left( \sqrt{g^{rr}} \; \frac{\partial
\sqrt{g_{tt}}}{\partial r}\right) \right|_{r = r_H} = \frac1{2 \pi
\rho_0^2}(r - r_H)|_{r = r_H} = 0.
\end{equation}
To calculate the entropy we apply Sen's formula \cite{Sen:2005wa,
Sen:2005iz} appealing to the near horizon data. Using
(\ref{expanElec}) we can write the near horizon solution  as
\begin{eqnarray}
ds^2 &=& - \frac{(r - r_H)^2}{\rho_0^2} dt^2 + \frac{\rho_0^2}{(r
- r_H)^2} dr^2 + \rho_0^2 d\Omega_2^2,
\nonumber \\
\phi &=& \frac1{2 a} \ln \frac{\rho_0^2}{4 \alpha}, \qquad F_{[2]}
= \frac{2 \sqrt \alpha}{\rho_0^2} \; dt \wedge dr.
\end{eqnarray}
To apply Sen's formula \cite{Sen:2005wa, Sen:2005iz}, we rewrite
it as follows
\begin{eqnarray} \label{bnh}
ds^2 &=& v_1 \left( - \hat r^2 d\hat t^2 + \frac{d \hat r^2}{\hat
r^2} \right) + v_2 d\Omega_2^2,
\nonumber \\
\phi &=& u, \qquad F_{\hat r \hat t} = e,
\end{eqnarray}
where $\hat r = r - r_H, \; \hat t = t / \rho_0^2$.

Now introduce the surface integrated Lagrangian density
\begin{equation} \label{Ld}
f(u, v_1, v_2, e) = \int d\theta d\phi \sqrt{g} \; L,
\end{equation}
and evaluate it using the near horizon data (\ref{bnh})
\begin{equation}
f(u, v_1, v_2, e) = \frac12 \left( v_1 - v_2 - 4 \alpha
\mathrm{e}^{2 a u} + e^2 \frac{v_2}{v_1} \mathrm{e}^{2 a u}
\right).
\end{equation}
The entropy function $F$ is the Legendre transform of this function
with respect to $e$:
\begin{equation}
q = \frac{\partial f}{\partial e} = e \frac{v_2}{v_1}
\mathrm{e}^{2au}, \qquad F = 2 \pi [ q e - f(u,v_1,v_2,e) ] = \pi
\left( v_2 - v_1 + 4 \alpha \mathrm{e}^{2 a u} + e^2
\frac{v_2}{v_1} \mathrm{e}^{2 a u} \right),
\end{equation}
or, in terms of $q$:
\begin{equation}
F = \pi \left( v_2 - v_1 + 4 \alpha \mathrm{e}^{2 a u} + q^2
\frac{v_1}{v_2} \mathrm{e}^{- 2 a u} \right).
\end{equation}
The entropy of the extremal black hole is given by the value of
the entropy function $F$ at   extremality:
\begin{equation} \label{extrF}
\frac{\partial F}{\partial u} = 0, \qquad \frac{\partial
F}{\partial v_1} = 0, \qquad \frac{\partial F}{\partial v_2} = 0.
\end{equation}
In our case, the extremality conditions (\ref{extrF}) read
\begin{equation}
\mathrm{e}^{- 2 a u} q^2 - v_2 = 0, \qquad - v_2^2 + \mathrm{e}^{- 2
a u} q^2 v_1 = 0, \qquad - \mathrm{e}^{-2 a u} q^2 v_1 + 4 \alpha
\mathrm{e}^{2 a u} v_2 = 0,
\end{equation}
leading to the solution
\begin{equation} \label{extrFs}
v_1 = 2 \sqrt\alpha q, \qquad v_2 = 2 \sqrt\alpha q, \qquad
\mathrm{e}^{2 a u} = \frac{q}{2 \sqrt\alpha}, \qquad (q > 0).
\end{equation}
Comparing with the local solution in our previous notation,
we get $q=q_e$.
Finally, substituting (\ref{extrFs}) in $F$ one
obtains the entropy
\begin{equation}
S = 4 \pi \sqrt\alpha \; q_e=2\pi\rho_0^2,
\end{equation}
which is precisely twice the Bekenstein-Hawking value. This is
similar to the result of refs. \cite{Dabholkar:2004yr,
Dabholkar:2004dq, Bak:2005mt}.

\section{Discussion}
In this paper, we have shown that in addition to charged black holes
with non-degenerate horizons, the DBG four-dimensional gravity
admits black hole solutions with the horizons of the $AdS_2 \times
S^2$ type. These solutions  form a one-parameter family  and exist
in a finite range  of the dilaton coupling constant $a$. New family
of solutions branch is disconnected from the branch of non-extremal
black holes which was studied earlier. Rather, it pinches off from
the extremal Reissner-Nordstr\"om black hole which {\em is} a
solution of the full EMDGB theory for $a=0$. Starting with zero $a$,
we were able to find global black hole solutions interpolating
between $AdS_2\times S^2$ at the horizon and Minkowski vacuum at
infinity for $a$ below some critical value which was found
numerically up to several decimals as $a_{\rm cr}\simeq
0.488219703$. Near the critical value $a\to a_{\rm cr}$, the mass
and the dilaton charge grow up, while their ratio saturates the BPS
bound of the EMD black holes. Similar feature  was observed  for the
charged  black holes in the Einstein-Born-Infeld-dilaton (EBID)
theory \cite{Clement:2000ue}.

It is worth noting that the family of electrically charged extremal
black holes in the EMDGB theory is  one-parametric  ($q_e$), while
the family of the corresponding extremal solutions  in the EMD
theory is two-parametric (with the parameters $q_e$ and
$\phi_\infty$). An asymptotic value of the dilaton is no more a free
parameter when the Gauss-Bonnet term is included, moreover, the
dilaton exponent $\e^{2a\phi_\infty}$ at the threshold $a=a_{\rm
cr}$ tends to zero for any finite value of the charge $q_e$.
Therefore,  modification of the extremal dilaton black hole by
higher curvature term consists not only in stretching its horizon to
a finite radius, but also in fixing the value of the dilaton at
infinity.

Our model can be viewed as a truncated heterotic string effective
theory in four dimensions. Whilst it does not include all quadratic
curvature terms,  it still exhibits features relevant to more
complete models, in particular, it predicts  correct entropy for
extremal black holes which  is twice the Bekenstein-Hawking entropy.
The dilaton varies from a finite value at the horizon to some
different finite value at infinity. The existence of the threshold
value of the dilaton coupling constant under which the global
solutions cease to exist is an interesting new phenomenon which may
be related to the string-black hole transition as described in
\cite{Cornalba:2006hc}\footnote{We thank Miguel Costa for bringing
the paper \cite{Cornalba:2006hc} to our attention.}. We think that
our model as well as the EBID model \cite{Clement:2000ue} (both
incorporating  typical stringy features) can be regarded as simple
toy models describing the string-black hole transition.

\section*{Acknowledgments}
The authors thank Soo-Jong Rey and Hideki Maeda for helpful
discussion and Miguel Costa for useful correspondence. DVG thanks
Department of Physics of NCU for hospitality and National Center of
Theoretical Sciences and (NCU) Center for Mathematics and Theoretic
Physics for support during his visit in January 2007. The work was
also supported in part by RFBR grant 02-04-16949. CMC and DGO were
supported by the National Science Council of the R.O.C. under the
grant NSC 95-2112-M-008-003. CMC was supported in part by National
Center of Theoretical Sciences and (NCU) Center for Mathematics and
Theoretic Physics.

\begin{appendix}
\section{Desingularization at the turning point}
Here we clarify the numerical procedure which allows us to continue
solutions through the singular  points. Rewrite the system
(\ref{Eqf1} - \ref{Eqphi}) as a matrix equation of the first order
\begin{equation}\label{dsys1}
A \frac{d}{dr}X = B,
\end{equation}
where $X$ is the six-dimensional vector consisting of the primary
dynamical variables $w, \rho, \mathrm{e}^{a \phi}$ and their first
derivatives with respect to the radial coordinate. The system
(\ref{dsys1}) has a regular solution provided $\det A \neq 0$. When
the solution approaches some point $r_{s}$ where
 $\det A \to 0$, the derivative $X'$ diverges as $O(1/{\det A})$.
In order to continue the solution through this point, we choose a
new independent variable $\sigma$ satisfying the condition
\begin{equation}
\dot r(\sigma) = \frac{dr}{d\sigma} \propto \det A.
\end{equation}
Then in terms of $\sigma$ the matrix equation (\ref{dsys1}) can be
rewritten in the regular form
\begin{equation} \label{dsys2}
A \dot X - B \dot r = 0.
\end{equation}
This desingularization is achieved by extending the set of unknown
functions to seven, considering the radial coordinate as a function
$r(\sigma)$. Denoting the seven-vector ($X(\sigma),\, r(\sigma)$) as
$Y(\sigma)$, one can see that the tangent vector has the unit
Euclidean metric norm:
\begin{equation}
\left| \frac{dY}{d\sigma} \right| = 1,
\end{equation}
provided the equation (\ref{dsys2}) holds:
\begin{equation}
|\dot r| = |\det A| \left( \det A^2 +  (X' \det A)^2
\right)^{-\frac12}.
\end{equation}
Using this desingularization one can continue the solution through
the singular turning point. This procedure is similar to one used in
\cite{Pomazanov:2000}. Geometrically this means gluing another
coordinate patch to the solution at singularity.

\end{appendix}



\begin{references}

\bibitem{Zwiebach:1985uq}
  B.~Zwiebach,
  ``Curvature squared terms and string theories,''
  Phys.\ Lett.\ B {\bf 156}, 315 (1985).

\bibitem{Callan:1986jb}
  C.~G.~Callan, I.~R.~Klebanov and M.~J.~Perry,
  ``String theory effective actions,''
  Nucl.\ Phys.\ B {\bf 278}, 78 (1986).

\bibitem{Gross:1986iv}
  D.~J.~Gross and E.~Witten,
  ``Superstring modifications of Einstein's equations,''
  Nucl.\ Phys.\ B {\bf 277}, 1 (1986).


\bibitem{Myers:1998gt}
  R.~C.~Myers,
  ``Black holes in higher curvature gravity,''
  arXiv:gr-qc/9811042.

\bibitem{Callan:1988hs}
  C.~G.~Callan, R.~C.~Myers and M.~J.~Perry,
  ``Black holes in string theory,''
  Nucl.\ Phys.\ B {\bf 311}, 673 (1989).



\bibitem{deWit:2005ya}
  B.~de Wit,
  ``Supersymmetric black holes,''
  Fortsch.\ Phys.\  {\bf 54}, 183 (2006)
  [arXiv:hep-th/0511261].

\bibitem{Mohaupt:2005jd}
  T.~Mohaupt,
  ``Strings, higher curvature corrections, and black holes,''
  arXiv:hep-th/0512048.


\bibitem{Wald:1993nt}
  R.~M.~Wald,
  ``Black hole entropy in the Noether charge,''
  Phys.\ Rev.\ D {\bf 48}, 3427 (1993)
  [arXiv:gr-qc/9307038].

\bibitem{Jacobson:1993vj}
  T.~Jacobson, G.~Kang and R.~C.~Myers,
  ``On black hole entropy,''
  Phys.\ Rev.\ D {\bf 49}, 6587 (1994)
  [arXiv:gr-qc/9312023].

\bibitem{Iyer:1994ys}
  V.~Iyer and R.~M.~Wald,
   ``Some properties of Noether charge and a proposal for dynamical black hole
  entropy,''
  Phys.\ Rev.\ D {\bf 50}, 846 (1994)
  [arXiv:gr-qc/9403028].

\bibitem{Jacobson:1994qe}
  T.~Jacobson, G.~Kang and R.~C.~Myers,
  ``Black hole entropy in higher curvature gravity,''
  arXiv:gr-qc/9502009.


\bibitem{Behrndt:1998eq}
  K.~Behrndt, G.~Lopes Cardoso, B.~de Wit, D.~Lust, T.~Mohaupt and W.~A.~Sabra,
   ``Higher-order black-hole solutions in N = 2 supergravity and Calabi-Yau
  string backgrounds,''
  Phys.\ Lett.\ B {\bf 429}, 289 (1998)
  [arXiv:hep-th/9801081].

\bibitem{LopesCardoso:1998wt}
  G.~Lopes Cardoso, B.~de Wit and T.~Mohaupt,
  ``Corrections to macroscopic supersymmetric black-hole entropy,''
  Phys.\ Lett.\ B {\bf 451}, 309 (1999)
  [arXiv:hep-th/9812082].

\bibitem{LopesCardoso:1999cv}
  G.~Lopes Cardoso, B.~de Wit and T.~Mohaupt,
  ``Deviations from the area law for supersymmetric black holes,''
  Fortsch.\ Phys.\  {\bf 48}, 49 (2000)
  [arXiv:hep-th/9904005].

\bibitem{LopesCardoso:1999ur}
  G.~Lopes Cardoso, B.~de Wit and T.~Mohaupt,
   ``Macroscopic entropy formulae and non-holomorphic corrections for
  supersymmetric black holes,''
  Nucl.\ Phys.\ B {\bf 567}, 87 (2000)
  [arXiv:hep-th/9906094].

\bibitem{LopesCardoso:1999xn}
  G.~Lopes Cardoso, B.~de Wit and T.~Mohaupt,
  ``Area law corrections from state counting and supergravity,''
  Class.\ Quant.\ Grav.\  {\bf 17}, 1007 (2000)
  [arXiv:hep-th/9910179].

\bibitem{Mohaupt:2000mj}
  T.~Mohaupt,
  ``Black hole entropy, special geometry and strings,''
  Fortsch.\ Phys.\  {\bf 49}, 3 (2001)
  [arXiv:hep-th/0007195].


\bibitem{LopesCardoso:2000qm}
  G.~Lopes Cardoso, B.~de Wit, J.~Kappeli and T.~Mohaupt,
  ``Stationary BPS solutions in N = 2 supergravity with R**2 interactions,''
  JHEP {\bf 0012}, 019 (2000)
  [arXiv:hep-th/0009234].


\bibitem{LopesCardoso:2000fp}
  G.~Lopes Cardoso, B.~de Wit, J.~Kappeli and T.~Mohaupt,
  ``Examples of stationary BPS solutions in N = 2 supergravity theories with
  R**2-interactions,''
  Fortsch.\ Phys.\  {\bf 49}, 557 (2001)
  [arXiv:hep-th/0012232].


\bibitem{Dabholkar:2004yr}
  A.~Dabholkar,
  ``Exact counting of black hole microstates,''
  Phys.\ Rev.\ Lett.\  {\bf 94}, 241301 (2005)
  [arXiv:hep-th/0409148].


\bibitem{Dabholkar:2004dq}
  A.~Dabholkar, R.~Kallosh and A.~Maloney,
  ``A stringy cloak for a classical singularity,''
  JHEP {\bf 0412}, 059 (2004)
  [arXiv:hep-th/0410076].

\bibitem{Sen:2004dp}
  A.~Sen,
  ``How does a fundamental string stretch its horizon?,''
  JHEP {\bf 0505}, 059 (2005)
  [arXiv:hep-th/0411255].

\bibitem{Hubeny:2004ji}
  V.~Hubeny, A.~Maloney and M.~Rangamani,
  ``String-corrected black holes,''
  JHEP {\bf 0505}, 035 (2005)
  [arXiv:hep-th/0411272].

\bibitem{Bak:2005mt}
  D.~Bak, S.~Kim and S.~J.~Rey,
  ``Exactly soluble BPS black holes in higher curvature N = 2 supergravity,''
  arXiv:hep-th/0501014.


\bibitem{Goldstein:2005hq}
  K.~Goldstein, N.~Iizuka, R.~P.~Jena and S.~P.~Trivedi,
  ``Non-supersymmetric attractors,''
  Phys.\ Rev.\ D {\bf 72}, 124021 (2005)
  [arXiv:hep-th/0507096].

\bibitem{Kallosh:2005ax}
  R.~Kallosh,
  ``New attractors,''
  JHEP {\bf 0512}, 022 (2005)
  [arXiv:hep-th/0510024].

\bibitem{Tripathy:2005qp}
  P.~K.~Tripathy and S.~P.~Trivedi,
  ``Non-supersymmetric attractors in string theory,''
  JHEP {\bf 0603}, 022 (2006)
  [arXiv:hep-th/0511117].

\bibitem{Giryavets:2005nf}
  A.~Giryavets,
  ``New attractors and area codes,''
  JHEP {\bf 0603}, 020 (2006)
  [arXiv:hep-th/0511215].

\bibitem{Goldstein:2005rr}
  K.~Goldstein, R.~P.~Jena, G.~Mandal and S.~P.~Trivedi,
  ``A C-function for non-supersymmetric attractors,''
  JHEP {\bf 0602}, 053 (2006)
  [arXiv:hep-th/0512138].

\bibitem{Kallosh:2006bt}
  R.~Kallosh, N.~Sivanandam and M.~Soroush,
  ``The non-BPS black hole attractor equation,''
  JHEP {\bf 0603}, 060 (2006)
  [arXiv:hep-th/0602005].

\bibitem{Kallosh:2006bx}
  R.~Kallosh,
  ``From BPS to non-BPS black holes canonically,''
  arXiv:hep-th/0603003.

\bibitem{Prester:2005qs}
  P.~Prester,
  ``Lovelock type gravity and small black holes in heterotic string theory,''
  JHEP {\bf 0602}, 039 (2006)
  [arXiv:hep-th/0511306].

\bibitem{Alishahiha:2006ke}
  M.~Alishahiha and H.~Ebrahim,
  ``Non-supersymmetric attractors and entropy function,''
  JHEP {\bf 0603}, 003 (2006)
  [arXiv:hep-th/0601016].

\bibitem{Sinha:2006yy}
  A.~Sinha and N.~V.~Suryanarayana,
  ``Extremal single-charge small black holes: Entropy function analysis,''
  Class.\ Quant.\ Grav.\  {\bf 23}, 3305 (2006)
  [arXiv:hep-th/0601183].

\bibitem{Chandrasekhar:2006kx}
  B.~Chandrasekhar, S.~Parvizi, A.~Tavanfar and H.~Yavartanoo,
  ``Non-supersymmetric attractors in R**2 gravities,''
  JHEP {\bf 0608}, 004 (2006)
  [arXiv:hep-th/0602022].

\bibitem{Parvizi:2006uz}
  S.~Parvizi and A.~Tavanfar,
   ``Partition function of non-supersymmetric black holes in the supergravity
  limit,''
  arXiv:hep-th/0602292.

\bibitem{Sahoo:2006rp}
  B.~Sahoo and A.~Sen,
   ``Higher derivative corrections to non-supersymmetric extremal black holes in
  N = 2 supergravity,''
  JHEP {\bf 0609}, 029 (2006)
  [arXiv:hep-th/0603149].


\bibitem{Astefanesei:2006sy}
  D.~Astefanesei, K.~Goldstein and S.~Mahapatra,
   ``Moduli and (un)attractor black hole thermodynamics,''
  arXiv:hep-th/0611140.
\bibitem{Sen:2005wa}
  A.~Sen,
   ``Black hole entropy function and the attractor mechanism in higher
  derivative gravity,''
  JHEP {\bf 0509}, 038 (2005)
  [arXiv:hep-th/0506177].


\bibitem{Sen:2005iz}
  A.~Sen,
  ``Entropy function for heterotic black holes,''
  JHEP {\bf 0603}, 008 (2006)
  [arXiv:hep-th/0508042].


\bibitem{Mignemi:1992nt}
  S.~Mignemi and N.~R.~Stewart,
  ``Charged black holes in effective string theory,''
  Phys.\ Rev.\ D {\bf 47}, 5259 (1993)
  [arXiv:hep-th/9212146].


\bibitem{Mignemi:1993ce}
  S.~Mignemi,
  ``Dyonic black holes in effective string theory,''
  Phys.\ Rev.\ D {\bf 51}, 934 (1995)
  [arXiv:hep-th/9303102].



\bibitem{Kanti:1995vq}
  P.~Kanti, N.~E.~Mavromatos, J.~Rizos, K.~Tamvakis and E.~Winstanley,
  ``Dilatonic Black Holes in Higher Curvature String Gravity,''
  Phys.\ Rev.\ D {\bf 54}, 5049 (1996)
  [arXiv:hep-th/9511071].

\bibitem{Torii:1996yi}
  T.~Torii, H.~Yajima and K.~I.~Maeda,
  ``Dilatonic black holes with Gauss-Bonnet term,''
  Phys.\ Rev.\ D {\bf 55}, 739 (1997)
  [arXiv:gr-qc/9606034].

\bibitem{Alexeev:1996vs}
  S.~O.~Alexeev and M.~V.~Pomazanov,
  ``Black hole solutions with dilatonic hair in higher curvature gravity,''
  Phys.\ Rev.\ D {\bf 55}, 2110 (1997)
  [arXiv:hep-th/9605106].


\bibitem{Alexeev:1997ua}
  S.~O.~Alexeev and M.~V.~Pomazanov,
  ``Singular regions in black hole solutions in higher order curvature
  gravity,''
  arXiv:gr-qc/9706066.


\bibitem{Melis:2005xt}
  M.~Melis and S.~Mignemi,
  ``Global properties of dilatonic Gauss-Bonnet black holes,''
  Class.\ Quant.\ Grav.\  {\bf 22}, 3169 (2005)
  [arXiv:gr-qc/0501087].


\bibitem{Melis:2005ji}
  M.~Melis and S.~Mignemi,
  ``Global properties of charged dilatonic Gauss-Bonnet black holes,''
  Phys.\ Rev.\ D {\bf 73}, 083010 (2006)
  [arXiv:gr-qc/0512132].

\bibitem{Mignemi:2006}
   S.~Mignemi,
  ``Black hole solutions of dimensionally reduced
  Einstein-Gauss-Bonnet gravity,''
  [arXiv:gr-qc/0607005].

\bibitem{Melis:2006}
  M.~Melis and  S.~Mignemi,
  ``Black hole solutions of dimensionally reduced
  Einstein-Gauss-Bonnet gravity with a cosmological constant,''
  [arXiv:gr-qc/0609133].


\bibitem{Torii:1998gm}
  T.~Torii and K.~I.~Maeda,
  ``Stability of a dilatonic black hole with a Gauss-Bonnet term,''
  Phys.\ Rev.\ D {\bf 58}, 084004 (1998).

\bibitem{Dotti:2004sh}
  G.~Dotti and R.~J.~Gleiser,
   ``Gravitational instability of Einstein-Gauss-Bonnet black holes under
  tensor mode perturbations,''
  Class.\ Quant.\ Grav.\  {\bf 22}, L1 (2005)
  [arXiv:gr-qc/0409005].


\bibitem{Dotti:2005sq}
  G.~Dotti and R.~J.~Gleiser,
   ``Linear stability of Einstein-Gauss-Bonnet static spacetimes. part. I:
  Tensor perturbations,''
  Phys.\ Rev.\ D {\bf 72}, 044018 (2005)
  [arXiv:gr-qc/0503117].

\bibitem{Gleiser:2005ra}
  R.~J.~Gleiser and G.~Dotti,
   ``Linear stability of Einstein-Gauss-Bonnet static spacetimes. II: Vector
  and scalar perturbations,''
  Phys.\ Rev.\ D {\bf 72}, 124002 (2005)
  [arXiv:gr-qc/0510069].


\bibitem{Moura:2006pz}
  F.~Moura and R.~Schiappa,
   ``Higher-derivative corrected black holes: Perturbative stability and
  absorption cross-section in heterotic string theory,''
  arXiv:hep-th/0605001.




\bibitem{Gibbons:1982ih}
  G.~W.~Gibbons,
  ``Antigravitating black hole solitons with scalar hair in N=4 supergravity,''
  Nucl.\ Phys.\ B {\bf 207} (1982) 337.


\bibitem{Gibbons:1987ps}
  G.~W.~Gibbons and K.~I.~Maeda,
  ``Black holes and membranes in higher dimensional theories with dilaton
  fields,''
  Nucl.\ Phys.\ B {\bf 298}, 741 (1988).

\bibitem{Gibbons:1985ac}
  G.~W.~Gibbons and D.~L.~Wiltshire,
  ``Black holes In Kaluza-Klein theory,''
  Annals Phys.\  {\bf 167}, 201 (1986)
  [Erratum-ibid.\  {\bf 176}, 393 (1987)].

\bibitem{Garfinkle:1990qj}
  D.~Garfinkle, G.~T.~Horowitz and A.~Strominger,
  ``Charged black holes in string theory,''
  Phys.\ Rev.\ D {\bf 43}, 3140 (1991)
  [Erratum-ibid.\ D {\bf 45}, 3888 (1992)].

\bibitem{Donets:1995ya}
  E.~E.~Donets and D.~V.~Gal'tsov,
  ``Stringy sphalerons and Gauss-Bonnet term,''
  Phys.\ Lett.\ B {\bf 352}, 261 (1995)
  [arXiv:hep-th/9503092].


\bibitem{Clement:2000ue}
  G.~Clement and D.~Gal'tsov,
   ``Solitons and black holes in Einstein-Born-Infeld-dilaton theory,''
  Phys.\ Rev.\ D {\bf 62}, 124013 (2000)
  [arXiv:hep-th/0007228].


\bibitem{Cornalba:2006hc}
  L.~Cornalba, M.~S.~Costa, J.~Penedones and P.~Vieira,
  ``From fundamental strings to small black holes,''
  JHEP {\bf 0612} (2006) 023
  [arXiv:hep-th/0607083].


\bibitem{Pomazanov:2000}
M.~V.~Pomazanov, ``On the structure of some typical singularities
for implicit ordinary differential equations,''
[arXiv:math-ph/0007008].


\end{references}
\end{document}